\documentclass[prx,aps,amsmath,amssymb,
twocolumn,
floatfix,
superscriptaddress]{revtex4-1}
\usepackage{amsthm}
\usepackage{amsfonts}
\usepackage{siunitx}
\usepackage{amsmath}
\usepackage{amssymb}
\usepackage{graphicx}
\usepackage{verbatim}
\usepackage[colorlinks]{hyperref}
\usepackage{tikz}
\usepackage{pgfplots}
\usepackage{braket}

\definecolor{linkcolor}{RGB}{0,83,166}
\hypersetup{colorlinks = true,allcolors = {linkcolor}}

\begin{document}
\title{Quantum annealing simulation of out-of-equilibrium magnetization in a spin-chain compound}
\author{Andrew D.~King}\email[]{aking@dwavesys.com}
\affiliation{D-Wave, Burnaby, BC, Canada V5G 4M9}
\author{Cristian D.~Batista}
\affiliation{Department of Physics and Astronomy, University of Tennessee, Knoxville, Tennessee 37996-1200, USA}
\affiliation{Neutron Scattering Division and Shull-Wollan Center, Oak Ridge National Laboratory, Oak Ridge, Tennessee 37831, USA}
\author{Jack Raymond}
\affiliation{D-Wave, Burnaby, BC, Canada V5G 4M9}
\author{Trevor Lanting}
\affiliation{D-Wave, Burnaby, BC, Canada V5G 4M9}
\author{Isil Ozfidan}
\affiliation{D-Wave, Burnaby, BC, Canada V5G 4M9}
\author{Gabriel Poulin-Lamarre}
\affiliation{D-Wave, Burnaby, BC, Canada V5G 4M9}
\author{Hao Zhang}
\affiliation{Department of Physics and Astronomy, University of Tennessee, Knoxville, Tennessee 37996-1200, USA}
\affiliation{Materials Science and Technology Division, Oak Ridge National Laboratory, Oak Ridge, Tennessee 37831, USA}
\author{Mohammad H.~Amin}
\affiliation{D-Wave, Burnaby, BC, Canada V5G 4M9}
\affiliation{Department of Physics, Simon Fraser University, Burnaby, BC, Canada V5A 1S6}

\date{\today}

\newcommand{\cco}{$\textrm{Ca}_3\textrm{Co}_2\textrm{O}_6$}
\newcommand{\tmgo}{$\textrm{TmMgGaO}_4$}
\begin{abstract}
  Geometrically frustrated spin-chain compounds such as \cco\ exhibit extremely slow relaxation under a changing magnetic field.  Consequently, both low-temperature laboratory experiments and Monte Carlo simulations have shown peculiar out-of-equilibrium magnetization curves, which arise from trapping in metastable configurations.  In this work we simulate this phenomenon in a superconducting quantum annealing processor, allowing us to probe the impact of quantum fluctuations on both equilibrium and dynamics of the system.  Increasing the quantum fluctuations with a transverse field reduces the impact of metastable traps in out-of-equilibrium samples, and aids the development of three-sublattice ferrimagnetic (up-up-down) long-range order.  At equilibrium we identify a finite-temperature shoulder in the 1/3-to-saturated phase transition, promoted by quantum fluctuations but with entropic origin.  This work demonstrates the viability of dynamical as well as equilibrium studies of frustrated magnetism using large-scale programmable quantum systems, and is therefore an important step toward programmable simulation of dynamics in materials using quantum hardware.
\end{abstract}

\maketitle

\section{Introduction}

Geometrically frustrated magnetic systems exhibit a variety of interesting dynamical and equilibrium properties.  The calcium oxalate \cco, a canonical example of a geometrically frustrated spin-chain compound, can be extremely slow to relax from metastable configurations when subjected to a changing magnetic field.  The magnetic $\textrm{Co}^{3+}$ ions align in ferromagnetic (FM) spin chains along the $c$ axis that form a triangular antiferromagnetic lattice in the $a$-$b$ plane.  This compound shows unusual out-of-equilibrium magnetization curves, a phenomenon which has been studied both  {\em in situ} \cite{Hardy2004a,Kim2018,Agrestini2008,Cheng2009,Fleck2010} and in simulation with a variety of simulated 2D and 3D models \cite{Kamiya2012,Kudasov2010,Kudasov2008,Soto2009}.  Hysteresis in this system shows counterintuitive response to changes in a longitudinal field: increasing the field sweep rate can increase the response to the changing field, as measured by bulk magnetization.  Simulations implicate the FM spin chains in the slow dynamics, and indicate the importance of intra-chain physics in correctly understanding the magnetic behavior of \cco\ as a rare example of effective low-dimensional Ising-like triangular frustration.  Some models have suggested quantum tunneling within the FM spin chains \cite{Maignan2004}; the influence of quantum fluctuations on relaxation in spin-chain compounds is therefore an important question.

\begin{figure}
  \includegraphics[height=6cm]{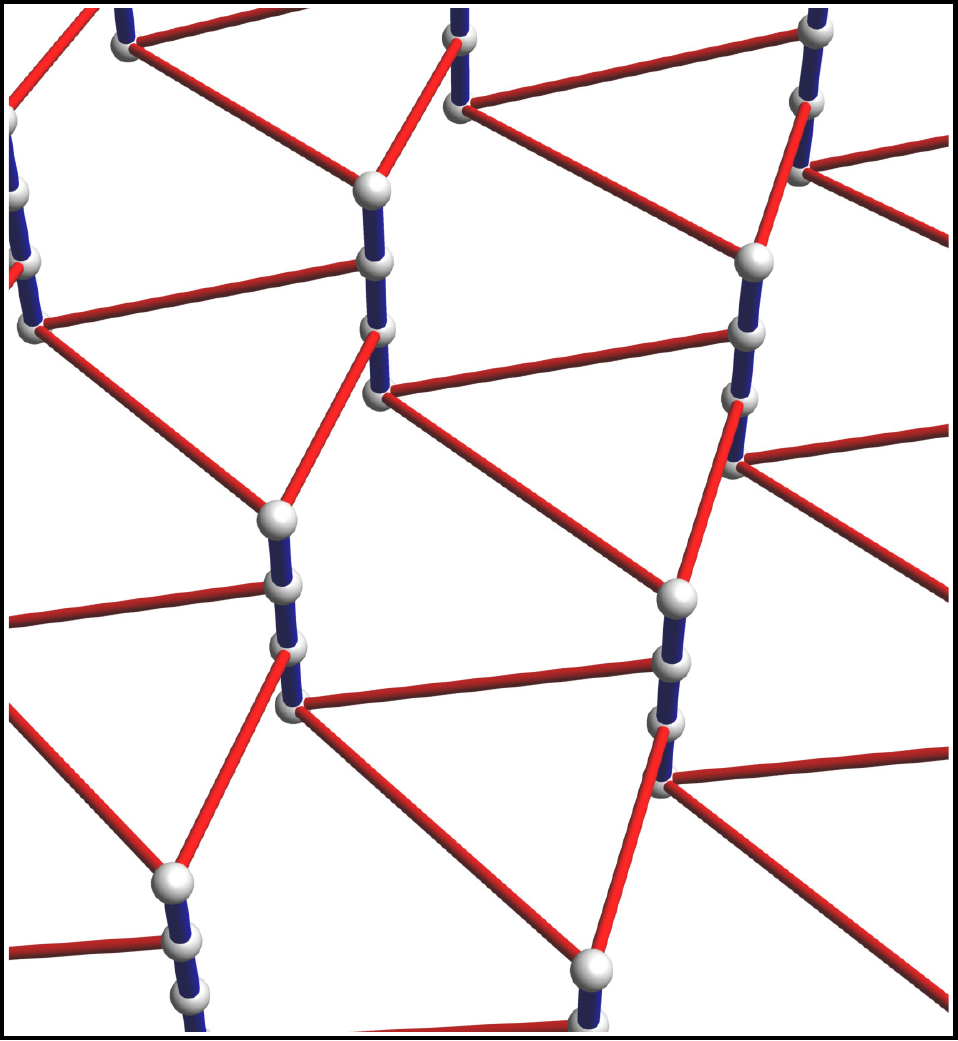}\\
  \caption{{\bf Geometrically frustrated square-octagonal lattice viewed as a spin-chain antiferromagnet.}  Red and blue lines indicate antiferromagnetic and ferromagnetic Ising exchange with couplings $J_1$ and $J_0=-1.8J_1$ respectively, where ferromagnetic bonds form four-spin chains.  As in \cco, the strong FM chains interact via weaker AFM couplings in a triangular geometry.}
  \label{fig:1}
\end{figure}

Two decades ago, Brooke et al.~explored the role of quantum fluctuations for the disordered spin glass LiHo$_x$Y$_{1-x}$F$_4$, annealing it to a low-energy state in a laboratory by attenuating thermal and quantum fluctuations \cite{Brooke1999}.  Their finding that quantum annealing relaxed the system faster than thermal annealing was a major motivating factor in the development of programmable superconducting quantum annealing (QA) processors \cite{Johnson2011}.  In turn, recent experiments have demonstrated that a class of quantum condensed matter systems can be simulated using QA processors \cite{Harris2018,King2018,King2019,Weinberg2019,Zhou2020}.  One such system is a geometrically-frustrated 2D Ising magnet (Fig.~\ref{fig:1}), in which quantum fluctuations induce order-by-disorder (OBD) at zero longitudinal field.  Upon increasing temperature, this ordering is suppressed via a sequence of two Berezinskii-Kosterlitz-Thouless (BKT) transitions, a phenomenon previously demonstrated using path-integral quantum Monte Carlo (QMC) simulations~\cite{Isakov2003} and later using QA simulation to probe equilibrium and dynamical properties~\cite{King2018,King2019}.

\section{Spin-chain model and related compounds}

 In this work we consider the lattice studied in Refs.~\cite{King2018,King2019} as a spin-chain antiferromagnet (Fig.~\ref{fig:1}) and simulate it under a longitudinal magnetic field, in addition to the transverse field that induces quantum fluctuations.  The  Hamiltonian of the system is
\begin{multline}\label{eq:ham}
  \mathcal{H} =  B \sum_{j ; \nu\in\{1,2,3,4\}} \sigma^z_{j \nu} +  J_1 \sum_{\langle j, l \rangle} \sigma^z_{j \nu(j,l)} \sigma^z_{l \nu(l,j)}\\
- J_0  \sum_{j ;\nu\in\{1,2,3\}}  \sigma^z_{j \nu}  \sigma^z_{j \nu+1}  - \Gamma \sum_{j ; \nu\in\{1,2,3,4\}} \sigma^x_{j \nu},
\end{multline}
where $\nu\in\{1,2,3,4 \}$ is the spin index of each vertical chain, and $\nu(j, l)$ is the spin index of the  chain $j$ that interacts with a spin in the chain $l$. The intra-chain interaction is ferromagnetic $(J_0 >0)$ and the inter-chain interaction is antiferromagnetic ($J_1 >0$). The Hamiltonian parameters can be expressed as $B(t)= \mathcal J(t) H(t)$, $J_0(t)=\mathcal J(t) I_0$,
$J_1(t)=\mathcal J(t) I_1$ and $\Gamma(t)$ to indicate the time-dependent control over strengths of the Ising, longitudinal field, and transverse field terms respectively, where $I_0=1.8$, $I_1=1$ and $\mathcal J(t) \geq 0$. Note that the ratio between the FM and the AFM Ising interactions is time-independent $J_0/J_1=I_0/I_1=1.8$.  $H(t)$ ranges from $0$ to $2$.  Thus we can alternatively express the time-dependent Hamiltonian as
\begin{multline}\label{eq:ham_b}
  \mathcal{H} =  \mathcal J(t)\bigg(  H(t) \sum_{j ; \nu\in\{1,2,3,4\}} \sigma^z_{j \nu} +  \sum_{\langle j, l \rangle} \sigma^z_{j \nu(j,l)} \sigma^z_{l \nu(l,j)}\\
- 1.8 \sum_{j ;\nu\in\{1,2,3\}}  \sigma^z_{j \nu}  \sigma^z_{j \nu+1} \bigg) - \Gamma(t) \sum_{j ; \nu\in\{1,2,3,4\}} \sigma^x_{j \nu},
\end{multline}
where $\mathcal J(t)$ and $\Gamma(t)$ are the time-dependent Ising and transverse energy functions comprising the annealing schedule, familiar in the field of quantum annealing \cite{Harris2018}.  The choices of $I_0=1.8$ and $I_1=1$ ensure that---as in related materials of interest---spin chains have few breaks at equilibrium, and the lattice can be fully saturated with $H=2$.

While the spin Hamiltonian $\mathcal{H}$ cannot be directly applied to any real material, it can still be used to model the behavior of known quantum magnets.  In addition to \cco~\cite{Kageyama97a,Kageyama97b,Maignan00,Hardy04,Moyoshi11}, the compound \tmgo\ has recently become the subject of intense study~\cite{Cevallos18,LiHan20,Shen19,Liu20,LiYueshing20}.  Here the magnetic Tm$^{3+}$ ions form a perfect triangular lattice. These moments can be described in terms of an effective spin-1/2 variable because the  
two lowest crystal field (CF) levels are separated from the rest by an energy gap  that is much bigger than the Ising-like exchange interaction between different magnetic moments. Given that the Tm$^{3+}$ ion has total angular momentum $J=6$, there is no Kramers degeneracy and the two  lowest energy CF levels are singlets~\cite{Shen19,Liu20}. The energy splitting between  the two singlets corresponds to an {\it intrinsic}  transverse field acting on the spin-1/2 variables and the low-energy physics of Tm$^{3+}$ is described by a transverse field Ising model (TFIM) on a triangular lattice~\cite{Shen19,Liu20}, which coincides with the  effective low-energy model for $\mathcal{H}$ in the limit $\Gamma/J_0 \ll 1$. In this limit, the low-energy degrees of freedom of each chain are also spin-1/2 variables that interact via the effective Hamiltonian:
\begin{equation}
{\tilde   {\mathcal H} } =  
{\tilde B} \sum_j  {\tilde \sigma}^z_j +  {\tilde J}_1 \sum_{\langle j, l \rangle} {\tilde \sigma}^z_j  {\tilde \sigma}^z_l 
- {\tilde \Gamma} \sum_{j } {\tilde \sigma}^x_j,
\label{eq:hameff}
\end{equation}
where ${\tilde B} = 4 B$, ${\tilde J}_1=J_1$ and  ${\tilde \Gamma} = \frac {\Gamma^4}{ J^3_0} $.

\begin{figure}
  \includegraphics[width=8.5cm]{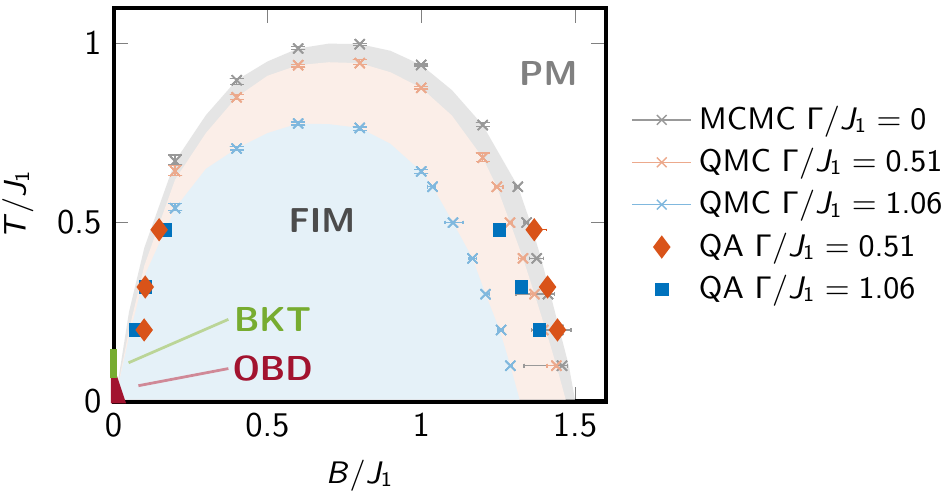}
  \caption{{\bf Thermodynamic phase diagram of $\mathcal{H}$ for different transverse fields $\Gamma$}.  For fixed value of $\Gamma/J_1$ the system has a ferrimagnetic (FIM) phase separated by a paramagnetic (PM) phase; phase transition locations are estimated using Monte Carlo and QA simulations.  For $\Gamma>0$, $B=0$ there are two BKT transitions delimiting a critical phase (BKT) and an ordered phase (OBD) \cite{Liu20,King2018}.}
  \label{fig:phase_diag}
\end{figure}

\section{Equilibrium properties}

As in \cite{King2018}, we program the QA processor to realize a cylindrical lattice on 1800 spins; in this experiment defects reduce the number of working spins to 1764.  To suppress boundary effects and better simulate the thermodynamic limit, we tune longitudinal field and coupling terms on a per-device basis to empirically homogenize qubit magnetizations and coupled spin-spin correlations, respectively \cite{Muller-Krumbhaar1972,King2018,Kairys2020}.

We estimate statistics of the system at thermal equilibrium under a fixed longitudinal magnetic field $B(t)=B(0)$ that ranges from $0$ to $2$, using the ``quantum evolution Monte Carlo'' (QEMC) method introduced in Ref.~\cite{King2018}.  In this mode of operation the system is initialized in a random classical spin state, and repeatedly exposed to both quantum and thermal fluctuations, in this case for $\SI{500}{\micro s}$ per exposure.  Between exposures, the fluctuations are quenched as the system is destructively projected to the computational $\sigma^z$ basis, and a classical state is read (see appendix).

Because the Ising system is repeatedly quenched and relaxed, the observed output does not precisely reflect the system at equilibrium.  Rather, prior to each readout the quantum and thermal fluctuations are turned off, and a small amount of local relaxation occurs during this process---most obviously in the erasure of single-spin excitations.  Bulk observables reflecting long-range order or magnetization are relatively protected from this distortion \cite{King2019}.  One can therefore think of the equilibrium estimates in two ways: first, as a perturbed observation of the system at equilibrium, or second, as an observation of a periodically-driven Floquet system.

Figure~\ref{fig:phase_diag} shows  the thermodynamic phase diagram obtained with classical ($\Gamma=0$) and quantum  ($\Gamma \neq 0$) Monte Carlo, and the transitions that are extracted from the QA processor for the QEMC runs.  In the classical Ising limit ($\Gamma = 0$), the phase diagram only  includes two phases.  The ordered phase at low enough temperature and longitudinal field $B \neq 0$ corresponds to a threefold degenerate three-sublattice up-up-down ferrimagnetic (FIM) state that induces a  1/3 plateau in the magnetization curve $M(B)/M_{\rm sat}$ ($M = \langle \sum_{j \nu} \sigma^z_{j \nu}\rangle$) and can be characterized by the  complex order parameter:
\begin{equation}
  \psi = me^{i\theta}= \big(m_1 + m_2e^{2\pi i/3} + m_3e^{4\pi i/3}\big)/\sqrt{3}
\end{equation}
where $m_1$, $m_2$, and $m_3$ are individual sublattice magnetizations \cite{Isakov2003}.  The three ground states of the $1/3$ plateau have $\psi$ equal to $-(2/\sqrt{3})e^{k2\pi i/3}$ for $k= 0,1,2$.  Thus to measure the onset of FIM long-range order (LRO) in our simulations of finite size lattices, we use the normalized cubic invariant
\begin{equation}
  m_{\text{FIM}}= 
- {\rm Re} [\big(\psi\sqrt{3}/2 \big)^3]
   = -m^3\cos(3\theta) \big(\sqrt{3}/2 \big)^3.
\end{equation}
The second phase is simply the paramagnetic (PM) state that becomes a fully polarized state along the longitudinal field direction at $T=0$.

As shown in Figure~\ref{fig:phase_diag}, the FIM phase survives for  $\Gamma \neq 0$. In addition,  there are two low-temperature phases identified in the $B=0$ case: a critical phase and an ordered phase, in which $\psi$ concentrates around six values rather than three~\cite{Moessner01}. This is also a three-sublattice ordering in which one of the two spin-up sublattices of the FIM state is polarized along the $x$-direction. The critical phase disappears for finite $B$ with the two BKT transitions that mark its upper and lower boundary.  
For $B\ll \Gamma$, the system transitions directly from the PM state to a sixfold degenerate state upon cooling, with the spins of the third sublattice canted along the longitudinal field direction.  Unfortunately, the base temperature of the QA processor is not low enough to access this ordered phase~\cite{King2018}.

\begin{figure}
  \includegraphics[scale = .8]{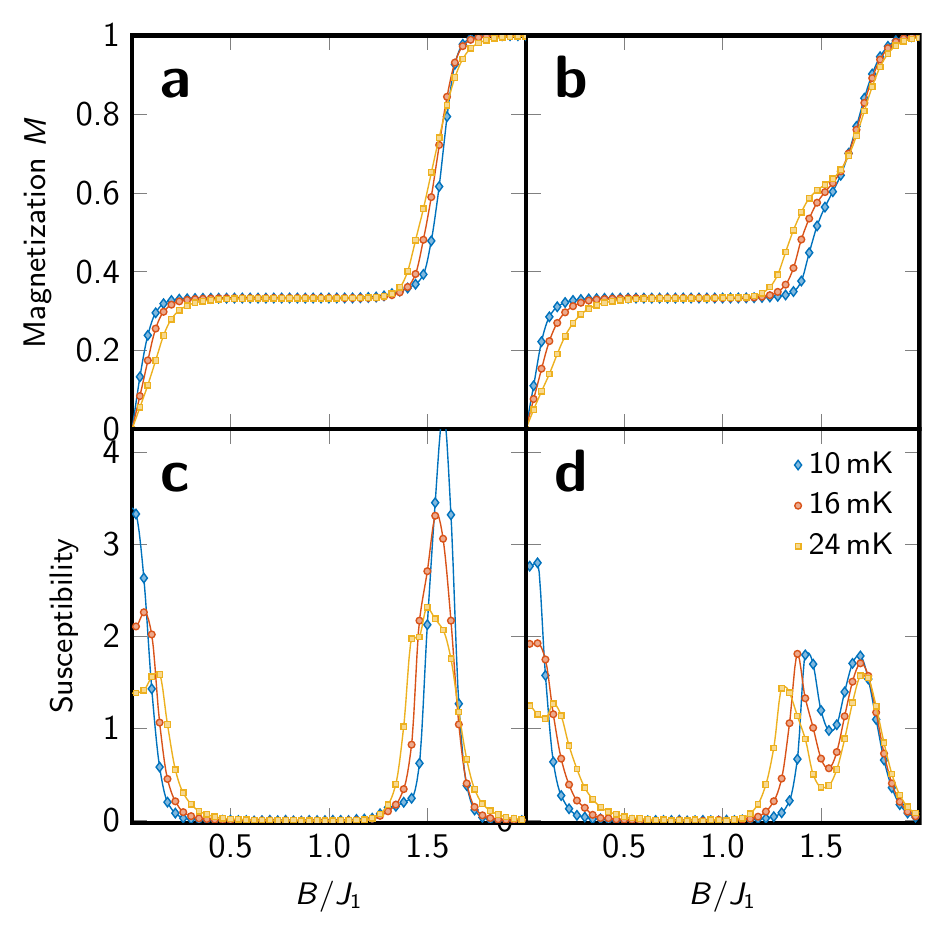}
  \caption{{\bf Equilibrium magnetization and susceptibility observations.} {\bf a}, {\bf c}: low tunneling, $\Gamma/J_1=0.51$.  {\bf b}, {\bf d}: high tunneling, $\Gamma/J_1=1.06$.  Susceptibility is computed as $dM/d(B/B_{\text{MAX}})$.  Increasing temperature corresponds to broadening of the transverse-field-induced magnetization shoulder, consistent with an entropic origin.}\label{fig:magsusc}
\end{figure}

The phase transition points between the FIM and the PM phases can be determined with the QA processor  (blue and red squares in Figure~\ref{fig:phase_diag}) by measuring the crossing point of $m_{\rm FIM}=1/2$.
The transition is  accompanied by a rather abrupt  change of the magnetization that produces a peak in the longitudinal susceptibility $dM/dB$. As shown in Fig.~\ref{fig:magsusc}, this peak splits into two upon increasing the temperature and the transverse field. The lower field peak still signals the metamagnetic phase transition between the FIM and the PM phases.
The higher field peak results from the combined effect of quantum and thermal fluctuations. 
As such, this second peak in $dM/dB$, or the corresponding magnetization shoulder in $M(B)$ (see Fig.~\ref{fig:magsusc}), can be used as an experimental fingerprint of the presence of a transverse field. In other words, the double-peak structure can be used to identify materials that are realizations of quantum Ising models.

\begin{figure*}
  \includegraphics[width=\linewidth]{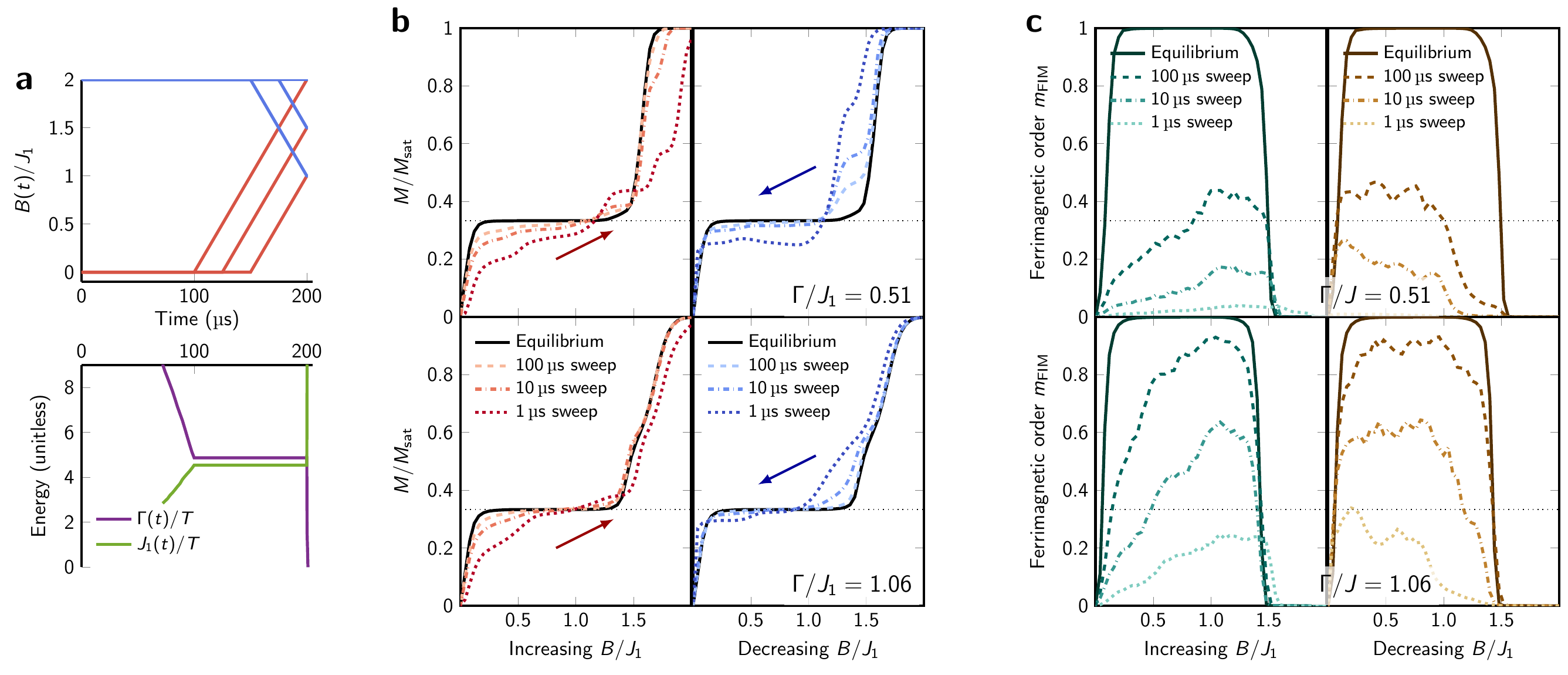}
  \caption{{\bf Out-of-equilibrium experimental results obtained from hysteresis simulation.}  {\bf a}, Annealing protocols for simulating hysteresis.  To take an observation at field $B_f$ from initial field $B_0$ of either zero or a saturating field $B_{\text{MAX}}=2J_1$, the system is annealed (via $\Gamma(t)$ and $J_1(t)$) under $B_0$; $B(t)$ is swept at the end of the protocol from $B_0$ to $B_f$ at a rate of $dt/d(B/B_{\text{MAX}}) =\SI{1}{\micro s}$, $\SI{10}{\micro s}$, or $\SI{100}{\micro s}$ before $J(t)$ and $\Gamma(t)$ are rapidly quenched prior to state readout ($B_f =  1$, $\tfrac 32$, and $2$ shown for $dt/d(B/B_{\text{MAX}}) =\SI{100}{\micro s}$).
    {\bf b}, Magnetization curves.  Shown are measurements at $T=\SI{10}{mK}$ for increasing field ($B_0=0$, left) and decreasing field ($B_0 = 2$), right) with low and high transverse field (top and bottom respectively).  Black curves show equilibrium estimates (Methods).
    {\bf c}, FIM long-range order $m_{\text{FIM}}$, which is saturated in the three-fold degenerate ground state of the $1/3$ plateau.
  }\label{fig:2}
\end{figure*}

To understand the origin of the second peak, it is convenient to start from the ${\tilde \Gamma}=0$  limit of ${\tilde   {\mathcal H} }$. In this classical limit, ${\tilde   {\mathcal H} }$ has an extensive ground state degeneracy  at the effective saturation field ${\tilde B}_{\rm sat} =  4 B_{\rm sat} = 6 J$, which is lifted by the transverse field: the ground space ${\cal S}$ is generated by spin states $| \phi_j \rangle$ that do not contain any pair of nearest-neighbor spins anti-aligned with the longitudinal field $B$. 
To first order in ${\tilde \Gamma}$, the new ground state is obtained by diagonalizing the ${\tilde \Gamma}$-term restricted to  ${\cal S}$. Given that the ${\tilde \Gamma}$-term is irreducible on ${\cal S}$ and all the off-diagonal terms are semi-negative defined, the Perron-Frobenius theorem guarantees that the transverse field  removes the ground state degeneracy {\it completely}. This observation implies that the new critical field for the  FIM-PM transition  is lower than ${\tilde B}_{\rm sat}=6 J$: ${\tilde B}_c({\tilde \Gamma},T=0) = 6 J - \delta {\tilde B}$ with $\delta {\tilde B} \simeq {\tilde \Gamma}$. This estimate results from requiring that the sum of the molecular and the applied field, $6J - {\tilde B}_c$ acting on an effective spin down of the FIM state must be comparable to ${\tilde \Gamma}$. As shown in Fig.~\ref{fig:magsusc}~(a), the discontinuous change of $M$ at $B=B_{\rm sat}$ and $\Gamma=0$ is replaced by a linear magnetization ramp that connects the 1/3 plateau ($M/M_{\rm sat}=1/3$) with the saturated value $M=M_{\rm sat}$.
The slope of this ramp is approximately $(1/M_{\rm sat}) \delta M/ \delta {\tilde B} \simeq 1/3 {\tilde \Gamma}$ .

The extensive ground state entropy reappears at a finite temperature $T \simeq \Gamma$ and induces a  magnetization shoulder near the middle of the magnetization ramp  that is clearly visible in  Fig.~\ref{fig:magsusc}~(b). A numerical estimate of the magnetization value that maximizes the  number of states $| \phi_j \rangle$  gives $M_m/M_{\rm sat} \simeq 0.583$, which is in good agreement with the value of the magnetization shoulder shown in Fig.~\ref{fig:magsusc}~(b). Further evidence for the entropic origin of the shoulder is provided in the appendix.
Based on the entropy argument, the width of the shoulder  is expected to be proportional to $T$ and to $\Gamma$ (note that $\delta {\tilde B} \simeq {\tilde \Gamma}$). In other words, the shoulder is the combined effect of quantum fluctuations, that stabilize the low energy manifold ${\cal S}$ over a finite longitudinal field interval of order $\Gamma$, and thermal fluctuations that select the sub-manifold of states with magnetization $M \simeq M_{\rm m}$. We note that an intermediate regime  between the FIM and the fully polarized phases has been detected with neutron scattering measurements of  \tmgo~\cite{LiYueshing20}. However, it is difficult to separate the relative roles of the intrinsic quantum and thermal fluctuations of the clean limit versus the spatial fluctuations induced by the significant amount of disorder present in \tmgo.

 \cco~comprises a triangular lattice
of ferromagnetic (FM) Ising chains coupled by weak antiferromagnetic (AFM) exchange interactions. This compound exhibits
longitudinal field-induced magnetization steps whose heights depend on the field sweep history and rate~\cite{Kageyama97a,Kageyama97b,Maignan00,Hardy04,Moyoshi11}.  These out-of-equilibrium magnetization steps, which appear at regular magnetic field intervals, originate from the extensive ground state degeneracy of the triangular Ising model and by the need to overcome an energy barrier to connect different ground states by  a sequence of individual spin flips (local dynamics). As it was pointed out in Ref.~\cite{Kudasov2006}, the regular spacing between the magnetization steps can be explained by the regular spacing between the molecular fields produced the six nearest-neighbor spins that surround the spin to be flipped.  Experimental studies of the quantum effect of an external transverse field in \cco~are challenging because of the very small effective gyromagnetic factor in the direction perpendicular to the chains.

\section{Out-of-equilibrium behavior}

We simulate relaxation of the system under field  $B(t)$ that either increases from $B(0)=0$ or decreases from a saturating field $B(0)=2J_1$.  Fig.~\ref{fig:2}a shows the time-dependent Hamiltonian terms for a $\SI{100}{\micro s}$ field sweep, which follows a $\SI{100}{\micro s}$ anneal to the desired values of $\Gamma$ and $J_1$ (cf.~\cite{Harris2018} Fig.~S13).  Since QA readout is achieved after projection to the $\sigma^z$ basis following a rapid quench of both $\Gamma$ and $J_1$, each value of $B/J_1$ must be simulated individually, as opposed to taking multiple measurements from a single sweep of $B$.  We probe sweep rates of $\SI{1}{\micro s}$, $\SI{10}{\micro s}$, and $\SI{100}{\micro s}$ for low and high transverse field ($\Gamma/J_1=0.51$ and $1.06$); $J_1 = \SI{1.89}{\giga\hertz}$ is roughly $4.5$ times the effective qubit temperature $T=\SI{10}{mK}$.  As the sweep of $B$ becomes slower, observations approach equilibrium values, which we estimate using a quantum evolution Monte Carlo (QEMC) protocol in which the system is relaxed iteratively \cite{King2018}, in contrast to the single-shot measurements.

Fig.~\ref{fig:2}b shows magnetization hysteresis curves for different sweep rates.  As in investigations of \cco, we observe distinct out-of-equilibrium behavior, including metastable overshooting of the equilibrium $1/3$ plateau.   Increasing the transverse field $\Gamma/J_1$ from $0.51$ to $1.06$ significantly reduces the signatures of metastability in the magnetization curves.

The $\SI{1}{\micro s}$ sweep of $B$ results in almost no LRO when $\Gamma/J_1=0.51$ despite the order seen at thermal equilibrium (Fig.~\ref{fig:2}c).  Throughout the curve, increasing $\Gamma/J_1$ to $1.06$ appears to hasten the development of LRO by over an order of magnitude.  In this sense it is clear that quantum fluctuations suppress metastable trapping in the out-of-equilibrium experiment.  Like in the case of \cco~\cite{Maignan2000}, the  out-of-equilibrium simulation of $\SI{1}{\micro s}$ sweep also shows three steps, in addition to the above-mentioned  shoulder.  It is important to note that  $\mathcal{H}$ is  far from being a realistic model for  \cco. Besides the fact that \cco~is a 3D material, a more realistic Hamiltonian should include multiple competing inter-chain exchange interactions that induce a commensurate-incommensurate transition at finite temperature~\cite{Kamiya2012}.  Nevertheless, as it was shown by Kudasov~\cite{Kudasov2006}, a simplified 2D triangular Ising model can explain the observation of three out-of-equilibrium steps, so it not surprising that a time-evolution controlled by $\mathcal{H}$ can reproduce the three steps.  Near-term QA processors will be able to simulate more flexible geometries \cite{Boothby2020}, including longer spin chains.

\begin{figure*}
  \includegraphics[scale = 0.8]{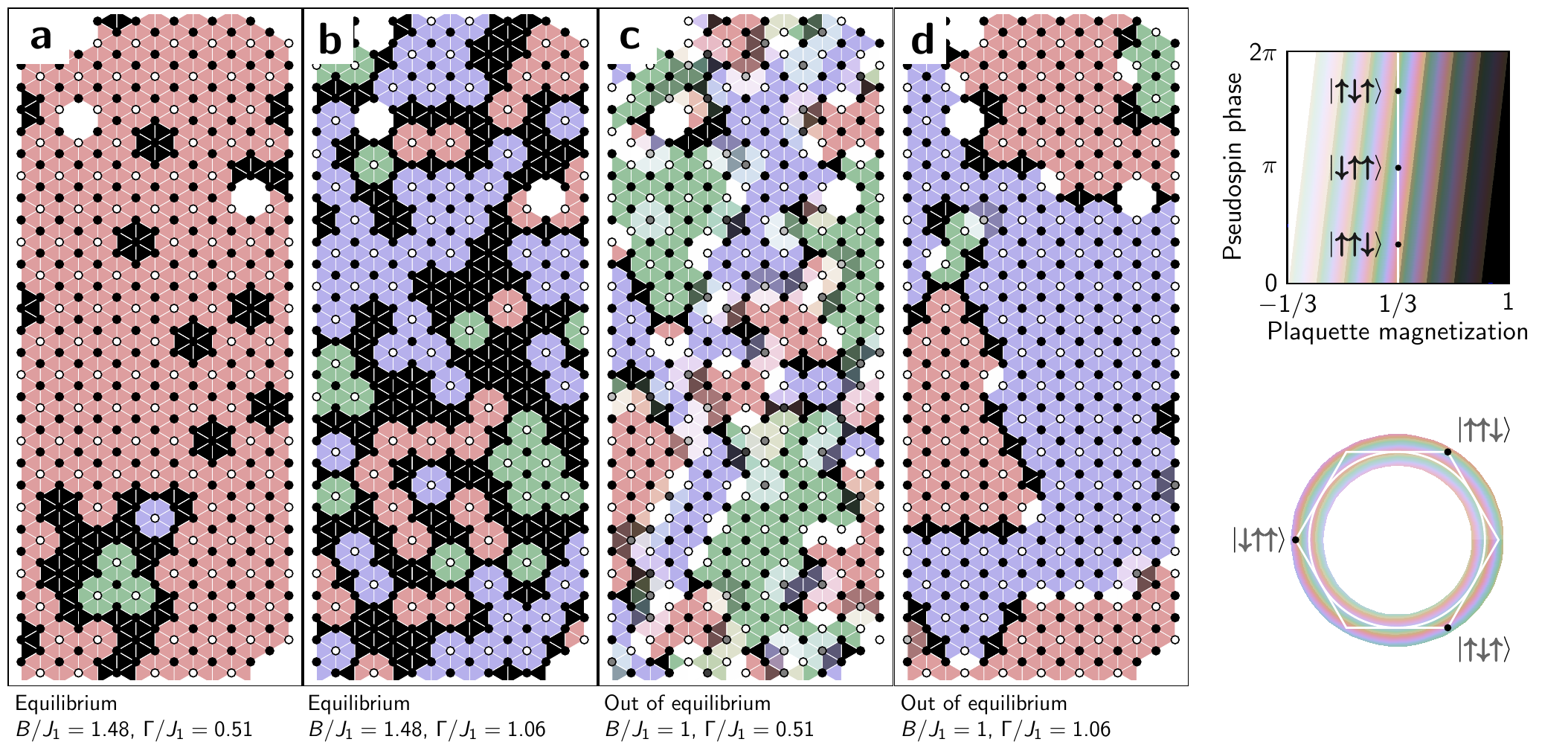}
  \caption{{\bf Experimental output samples.} Marker darkness indicates magnetization of the four-qubit chain; plaquette darkness indicates plaquette magnetization.  Plaquette hue indicates pseudospin phase; the 1/3 plateau has three degenerate FIM ground states with varying phase.  The lattice is periodic on the top/bottom and open at the left/right.  Six of 450 sites are empty due to inoperable qubits.  {\bf a--b}, Equilibrium samples at $H = 1.48$; the transverse field modulates long-range order.  {\bf a}, at $\Gamma/J_1=0.51$ the system adopts long-range FIM order and the system is dominated by plaquettes of a single phase, with $m_{\text{FIM}}$ large.  {\bf b}, with $\Gamma/J_1=1.06$, the system is characterized by competing FIM domains, with saturated plaquettes at the interfaces.  {\bf c--d}, out-of-equilibrium samples, with $H$ decreasing from $2$ to $1$ over $\SI{0.5}{\micro s}$.  {\bf c}, with $\Gamma/J_1=0.51$ the system is chaotic, with many small FIM domains.  The interfaces of these domains host trapped excitations, many in the form of broken four-qubit chains with magnetization zero.  {\bf d}, with $\Gamma/J_1=1.06$ the system forms large FIM domains during the sweep of $H$.  The interfaces between these domains host plaquettes of magnetization $+1$ and $-1/3$. }\label{fig:states}
\end{figure*}

The structure of output states can shed light on the effect of quantum fluctuations at equilibrium and out-of-equilibrium.  Fig.~\ref{fig:states} shows representative output states from the QA simulation. Fig.~\ref{fig:states}a and b show equilibrium states at $B/J_1=1.48$ for $\Gamma/J_1=0.51$ and $\Gamma/J_1=1.06$, respectively.  We see clear evidence of the transverse field's role in suppressing FIM LRO while remaining in the ${\cal S}$ manifold. In Fig.~\ref{fig:states}c and d we see out-of-equilibrium states for the fastest sweep of $B/J_1$ from $2$ to $1$.  In this case we see far more FIM LRO in the high-$\Gamma$ case (in line with Fig.~\ref{fig:2}c), and many competing FIM domains in the low-$\Gamma$ case.  At the domain boundaries we see a variety of trapped defects, including up-down-down plaquettes and  broken spin chains with nearly zero average magnetization,  that contribute to the overshooting of the $1/3$ plateau seen in Fig.~\ref{fig:2}b. The presence of broken chain defects indicates that the out of equilibirum dynamics is not fully captured by the low-energy effective model \cite{Kudasov2006}.

It is not clear to what extent this trapping is {\em escaped} versus {\em avoided} as the system passes through the phase transition either as $B$ increases from $0$ or decreases through $B_c\approx \tfrac 32J_1$.   Although path-integral Monte Carlo (PIMC) simulation is not expected to correctly reproduce open-system relaxation dynamics of complex frustrated systems \cite{Andriyash2017,King2019}, such simulations suggest that the acceleration arising from the transverse field is dynamical, and not reflective of a new path through the phase transition that avoids metastable configurations (see appendix).

\section{Conclusions}

We have demonstrated a large-scale out-of-equilibrium simulation of a frustrated transverse field Ising model using a programmable superconducting QA processor.  When quantum fluctuations are small, we observe the signature metastable magnetization curves also seen in experimental measurements of spin-chain compounds.  Introducing larger quantum fluctuations produces both dynamical and equilibrium effects on the system.  Out-of-equilibrium simulations are a promising application for quantum simulation: standard quantum Monte Carlo methods---be they single-spin Glauber dynamics or imaginary-time cluster update dynamics (e.g.~Swendsen-Wang)---cannot faithfully reproduce the  the real time-evolution of a quantum system \cite{Andriyash2017,Kechedzhi2018,King2019}.  Although evolution of a quantum system in the presence of a thermal bath is fundamental to QA, the relationship between QA dynamics and time-evolution of a quantum Ising model has not been established experimentally in detail for large systems.  Moreover, the spin-chain system we have simulated is only qualitatively similar to real compounds---future hardware generations with greater control and more flexible Hamiltonian terms and geometry \cite{Ozfidan2019,Boothby2020} will allow deeper and more realistic simulations of this nature.

Although independent control of transverse and longitudinal magnetic fields is difficult to achieve for {\em in situ} crystal studies, a laboratory reproduction of our findings in \cco\ would be of great interest.  This would mark the first example of a quantum simulation preceding confirmation in a physical sample, an important milestone in quantum simulation for materials science.

\section*{Acknowledgements}

We gratefully acknowledge the contributions of the teams at D-Wave who designed, fabricated, and calibrated the QA processor used in this work.  We thank Vivien Zapf, Arnab Banerjee, and Paul Kairys for helpful conversations.

\linespread{1}
\bibliography{paper_hysteresis}%

\begin{thebibliography}{39}%
\makeatletter
\providecommand \@ifxundefined [1]{%
 \@ifx{#1\undefined}
}%
\providecommand \@ifnum [1]{%
 \ifnum #1\expandafter \@firstoftwo
 \else \expandafter \@secondoftwo
 \fi
}%
\providecommand \@ifx [1]{%
 \ifx #1\expandafter \@firstoftwo
 \else \expandafter \@secondoftwo
 \fi
}%
\providecommand \natexlab [1]{#1}%
\providecommand \enquote  [1]{``#1''}%
\providecommand \bibnamefont  [1]{#1}%
\providecommand \bibfnamefont [1]{#1}%
\providecommand \citenamefont [1]{#1}%
\providecommand \href@noop [0]{\@secondoftwo}%
\providecommand \href [0]{\begingroup \@sanitize@url \@href}%
\providecommand \@href[1]{\@@startlink{#1}\@@href}%
\providecommand \@@href[1]{\endgroup#1\@@endlink}%
\providecommand \@sanitize@url [0]{\catcode `\\12\catcode `\$12\catcode
  `\&12\catcode `\#12\catcode `\^12\catcode `\_12\catcode `\%12\relax}%
\providecommand \@@startlink[1]{}%
\providecommand \@@endlink[0]{}%
\providecommand \url  [0]{\begingroup\@sanitize@url \@url }%
\providecommand \@url [1]{\endgroup\@href {#1}{\urlprefix }}%
\providecommand \urlprefix  [0]{URL }%
\providecommand \Eprint [0]{\href }%
\providecommand \doibase [0]{http://dx.doi.org/}%
\providecommand \selectlanguage [0]{\@gobble}%
\providecommand \bibinfo  [0]{\@secondoftwo}%
\providecommand \bibfield  [0]{\@secondoftwo}%
\providecommand \translation [1]{[#1]}%
\providecommand \BibitemOpen [0]{}%
\providecommand \bibitemStop [0]{}%
\providecommand \bibitemNoStop [0]{.\EOS\space}%
\providecommand \EOS [0]{\spacefactor3000\relax}%
\providecommand \BibitemShut  [1]{\csname bibitem#1\endcsname}%
\let\auto@bib@innerbib\@empty
\bibitem [{\citenamefont {Hardy}\ \emph
  {et~al.}(2004{\natexlab{a}})\citenamefont {Hardy}, \citenamefont {Flahaut},
  \citenamefont {Lees},\ and\ \citenamefont {Petrenko}}]{Hardy2004a}%
  \BibitemOpen
  \bibfield  {author} {\bibinfo {author} {\bibfnamefont {V.}~\bibnamefont
  {Hardy}}, \bibinfo {author} {\bibfnamefont {D.}~\bibnamefont {Flahaut}},
  \bibinfo {author} {\bibfnamefont {M.~R.}\ \bibnamefont {Lees}}, \ and\
  \bibinfo {author} {\bibfnamefont {O.~A.}\ \bibnamefont {Petrenko}},\ }\href
  {\doibase 10.1103/PhysRevB.70.214439} {\bibfield  {journal} {\bibinfo
  {journal} {Physical Review B}\ }\textbf {\bibinfo {volume} {70}},\ \bibinfo
  {pages} {1} (\bibinfo {year} {2004}{\natexlab{a}})}\BibitemShut {NoStop}%
\bibitem [{\citenamefont {Kim}\ \emph {et~al.}(2018)\citenamefont {Kim},
  \citenamefont {Mun}, \citenamefont {Ding}, \citenamefont {Hansen},
  \citenamefont {Jaime}, \citenamefont {Harrison}, \citenamefont {Yi},
  \citenamefont {Chai}, \citenamefont {Sun}, \citenamefont {Cheong},\ and\
  \citenamefont {Zapf}}]{Kim2018}%
  \BibitemOpen
  \bibfield  {author} {\bibinfo {author} {\bibfnamefont {J.~W.}\ \bibnamefont
  {Kim}}, \bibinfo {author} {\bibfnamefont {E.~D.}\ \bibnamefont {Mun}},
  \bibinfo {author} {\bibfnamefont {X.}~\bibnamefont {Ding}}, \bibinfo {author}
  {\bibfnamefont {A.}~\bibnamefont {Hansen}}, \bibinfo {author} {\bibfnamefont
  {M.}~\bibnamefont {Jaime}}, \bibinfo {author} {\bibfnamefont
  {N.}~\bibnamefont {Harrison}}, \bibinfo {author} {\bibfnamefont {H.~T.}\
  \bibnamefont {Yi}}, \bibinfo {author} {\bibfnamefont {Y.}~\bibnamefont
  {Chai}}, \bibinfo {author} {\bibfnamefont {Y.}~\bibnamefont {Sun}}, \bibinfo
  {author} {\bibfnamefont {S.~W.}\ \bibnamefont {Cheong}}, \ and\ \bibinfo
  {author} {\bibfnamefont {V.~S.}\ \bibnamefont {Zapf}},\ }\href {\doibase
  10.1103/PhysRevB.98.024407} {\bibfield  {journal} {\bibinfo  {journal}
  {Physical Review B}\ }\textbf {\bibinfo {volume} {98}},\ \bibinfo {pages} {1}
  (\bibinfo {year} {2018})}\BibitemShut {NoStop}%
\bibitem [{\citenamefont {Agrestini}\ \emph {et~al.}(2008)\citenamefont
  {Agrestini}, \citenamefont {Chapon}, \citenamefont {Daoud-Aladine},
  \citenamefont {Schefer}, \citenamefont {Gukasov}, \citenamefont {Mazzoli},
  \citenamefont {Lees},\ and\ \citenamefont {Petrenko}}]{Agrestini2008}%
  \BibitemOpen
  \bibfield  {author} {\bibinfo {author} {\bibfnamefont {S.}~\bibnamefont
  {Agrestini}}, \bibinfo {author} {\bibfnamefont {L.~C.}\ \bibnamefont
  {Chapon}}, \bibinfo {author} {\bibfnamefont {A.}~\bibnamefont
  {Daoud-Aladine}}, \bibinfo {author} {\bibfnamefont {J.}~\bibnamefont
  {Schefer}}, \bibinfo {author} {\bibfnamefont {A.}~\bibnamefont {Gukasov}},
  \bibinfo {author} {\bibfnamefont {C.}~\bibnamefont {Mazzoli}}, \bibinfo
  {author} {\bibfnamefont {M.~R.}\ \bibnamefont {Lees}}, \ and\ \bibinfo
  {author} {\bibfnamefont {O.~A.}\ \bibnamefont {Petrenko}},\ }\href {\doibase
  10.1103/PhysRevLett.101.097207} {\bibfield  {journal} {\bibinfo  {journal}
  {Physical Review Letters}\ }\textbf {\bibinfo {volume} {101}},\ \bibinfo
  {pages} {2} (\bibinfo {year} {2008})}\BibitemShut {NoStop}%
\bibitem [{\citenamefont {Cheng}\ \emph {et~al.}(2009)\citenamefont {Cheng},
  \citenamefont {Zhou},\ and\ \citenamefont {Goodenough}}]{Cheng2009}%
  \BibitemOpen
  \bibfield  {author} {\bibinfo {author} {\bibfnamefont {J.~G.}\ \bibnamefont
  {Cheng}}, \bibinfo {author} {\bibfnamefont {J.~S.}\ \bibnamefont {Zhou}}, \
  and\ \bibinfo {author} {\bibfnamefont {J.~B.}\ \bibnamefont {Goodenough}},\
  }\href {\doibase 10.1103/PhysRevB.79.184414} {\bibfield  {journal} {\bibinfo
  {journal} {Physical Review B - Condensed Matter and Materials Physics}\
  }\textbf {\bibinfo {volume} {79}},\ \bibinfo {pages} {1} (\bibinfo {year}
  {2009})}\BibitemShut {NoStop}%
\bibitem [{\citenamefont {Fleck}\ \emph {et~al.}(2010)\citenamefont {Fleck},
  \citenamefont {Lees}, \citenamefont {Agrestini}, \citenamefont {McIntyre},\
  and\ \citenamefont {Petrenko}}]{Fleck2010}%
  \BibitemOpen
  \bibfield  {author} {\bibinfo {author} {\bibfnamefont {C.~L.}\ \bibnamefont
  {Fleck}}, \bibinfo {author} {\bibfnamefont {M.~R.}\ \bibnamefont {Lees}},
  \bibinfo {author} {\bibfnamefont {S.}~\bibnamefont {Agrestini}}, \bibinfo
  {author} {\bibfnamefont {G.~J.}\ \bibnamefont {McIntyre}}, \ and\ \bibinfo
  {author} {\bibfnamefont {O.~A.}\ \bibnamefont {Petrenko}},\ }\href {\doibase
  10.1209/0295-5075/90/67006} {\bibfield  {journal} {\bibinfo  {journal} {EPL
  (Europhysics Letters)}\ }\textbf {\bibinfo {volume} {90}},\ \bibinfo {pages}
  {67006} (\bibinfo {year} {2010})}\BibitemShut {NoStop}%
\bibitem [{\citenamefont {Kamiya}\ and\ \citenamefont
  {Batista}(2012)}]{Kamiya2012}%
  \BibitemOpen
  \bibfield  {author} {\bibinfo {author} {\bibfnamefont {Y.}~\bibnamefont
  {Kamiya}}\ and\ \bibinfo {author} {\bibfnamefont {C.~D.}\ \bibnamefont
  {Batista}},\ }\href {\doibase 10.1103/PhysRevLett.109.067204} {\bibfield
  {journal} {\bibinfo  {journal} {Physical Review Letters}\ }\textbf {\bibinfo
  {volume} {109}},\ \bibinfo {pages} {2} (\bibinfo {year} {2012})}\BibitemShut
  {NoStop}%
\bibitem [{\citenamefont {Kudasov}\ \emph {et~al.}(2010)\citenamefont
  {Kudasov}, \citenamefont {Korshunov}, \citenamefont {Pavlov},\ and\
  \citenamefont {Maslov}}]{Kudasov2010}%
  \BibitemOpen
  \bibfield  {author} {\bibinfo {author} {\bibfnamefont {Y.~B.}\ \bibnamefont
  {Kudasov}}, \bibinfo {author} {\bibfnamefont {A.~S.}\ \bibnamefont
  {Korshunov}}, \bibinfo {author} {\bibfnamefont {V.~N.}\ \bibnamefont
  {Pavlov}}, \ and\ \bibinfo {author} {\bibfnamefont {D.~A.}\ \bibnamefont
  {Maslov}},\ }\href {\doibase 10.1007/s10909-009-0096-4} {\bibfield  {journal}
  {\bibinfo  {journal} {Journal of Low Temperature Physics}\ }\textbf {\bibinfo
  {volume} {159}},\ \bibinfo {pages} {76} (\bibinfo {year} {2010})}\BibitemShut
  {NoStop}%
\bibitem [{\citenamefont {Kudasov}\ \emph {et~al.}(2008)\citenamefont
  {Kudasov}, \citenamefont {Korshunov}, \citenamefont {Pavlov},\ and\
  \citenamefont {Maslov}}]{Kudasov2008}%
  \BibitemOpen
  \bibfield  {author} {\bibinfo {author} {\bibfnamefont {Y.~B.}\ \bibnamefont
  {Kudasov}}, \bibinfo {author} {\bibfnamefont {A.~S.}\ \bibnamefont
  {Korshunov}}, \bibinfo {author} {\bibfnamefont {V.~N.}\ \bibnamefont
  {Pavlov}}, \ and\ \bibinfo {author} {\bibfnamefont {D.~A.}\ \bibnamefont
  {Maslov}},\ }\href {\doibase 10.1103/PhysRevB.78.132407} {\bibfield
  {journal} {\bibinfo  {journal} {Physical Review B}\ }\textbf {\bibinfo
  {volume} {78}},\ \bibinfo {pages} {132407} (\bibinfo {year}
  {2008})}\BibitemShut {NoStop}%
\bibitem [{\citenamefont {Soto}\ \emph {et~al.}(2009)\citenamefont {Soto},
  \citenamefont {Mart{\'{i}}nez}, \citenamefont {Baibich}, \citenamefont
  {Florez},\ and\ \citenamefont {Vargas}}]{Soto2009}%
  \BibitemOpen
  \bibfield  {author} {\bibinfo {author} {\bibfnamefont {R.}~\bibnamefont
  {Soto}}, \bibinfo {author} {\bibfnamefont {G.}~\bibnamefont
  {Mart{\'{i}}nez}}, \bibinfo {author} {\bibfnamefont {M.~N.}\ \bibnamefont
  {Baibich}}, \bibinfo {author} {\bibfnamefont {J.~M.}\ \bibnamefont {Florez}},
  \ and\ \bibinfo {author} {\bibfnamefont {P.}~\bibnamefont {Vargas}},\ }\href
  {\doibase 10.1103/PhysRevB.79.184422} {\bibfield  {journal} {\bibinfo
  {journal} {Physical Review B - Condensed Matter and Materials Physics}\
  }\textbf {\bibinfo {volume} {79}},\ \bibinfo {pages} {1} (\bibinfo {year}
  {2009})}\BibitemShut {NoStop}%
\bibitem [{\citenamefont {Maignan}\ \emph {et~al.}(2004)\citenamefont
  {Maignan}, \citenamefont {Hardy}, \citenamefont {H{\'{e}}bert}, \citenamefont
  {Drillon}, \citenamefont {Lees}, \citenamefont {Petrenko}, \citenamefont
  {Paul},\ and\ \citenamefont {Khomskii}}]{Maignan2004}%
  \BibitemOpen
  \bibfield  {author} {\bibinfo {author} {\bibfnamefont {A.}~\bibnamefont
  {Maignan}}, \bibinfo {author} {\bibfnamefont {V.}~\bibnamefont {Hardy}},
  \bibinfo {author} {\bibfnamefont {S.}~\bibnamefont {H{\'{e}}bert}}, \bibinfo
  {author} {\bibfnamefont {M.}~\bibnamefont {Drillon}}, \bibinfo {author}
  {\bibfnamefont {M.~R.}\ \bibnamefont {Lees}}, \bibinfo {author}
  {\bibfnamefont {O.}~\bibnamefont {Petrenko}}, \bibinfo {author}
  {\bibfnamefont {D.~M.~K.}\ \bibnamefont {Paul}}, \ and\ \bibinfo {author}
  {\bibfnamefont {D.}~\bibnamefont {Khomskii}},\ }\href {\doibase
  10.1039/B316717H} {\bibfield  {journal} {\bibinfo  {journal} {J. Mater.
  Chem.}\ }\textbf {\bibinfo {volume} {14}},\ \bibinfo {pages} {1231} (\bibinfo
  {year} {2004})}\BibitemShut {NoStop}%
\bibitem [{\citenamefont {Brooke}\ \emph {et~al.}(1999)\citenamefont {Brooke},
  \citenamefont {Bitko}, \citenamefont {F.}, \citenamefont {Rosenbaum},\ and\
  \citenamefont {Aeppli}}]{Brooke1999}%
  \BibitemOpen
  \bibfield  {author} {\bibinfo {author} {\bibfnamefont {J.}~\bibnamefont
  {Brooke}}, \bibinfo {author} {\bibfnamefont {D.}~\bibnamefont {Bitko}},
  \bibinfo {author} {\bibfnamefont {T.}~\bibnamefont {F.}}, \bibinfo {author}
  {\bibnamefont {Rosenbaum}}, \ and\ \bibinfo {author} {\bibfnamefont
  {G.}~\bibnamefont {Aeppli}},\ }\href
  {http://science.sciencemag.org/content/284/5415/779.abstract} {\bibfield
  {journal} {\bibinfo  {journal} {Science}\ }\textbf {\bibinfo {volume}
  {284}},\ \bibinfo {pages} {779 LP } (\bibinfo {year} {1999})}\BibitemShut
  {NoStop}%
\bibitem [{\citenamefont {Johnson}\ \emph {et~al.}(2011)\citenamefont
  {Johnson}, \citenamefont {Amin}, \citenamefont {Gildert}, \citenamefont
  {Lanting}, \citenamefont {Hamze}, \citenamefont {Dickson}, \citenamefont
  {Harris}, \citenamefont {Berkley}, \citenamefont {Johansson}, \citenamefont
  {Bunyk}, \citenamefont {Chapple}, \citenamefont {Enderud}, \citenamefont
  {Hilton}, \citenamefont {Karimi}, \citenamefont {Ladizinsky}, \citenamefont
  {Ladizinsky}, \citenamefont {Oh}, \citenamefont {Perminov}, \citenamefont
  {Rich}, \citenamefont {Thom}, \citenamefont {Tolkacheva}, \citenamefont
  {Truncik}, \citenamefont {Uchaikin}, \citenamefont {Wang}, \citenamefont
  {Wilson},\ and\ \citenamefont {Rose}}]{Johnson2011}%
  \BibitemOpen
  \bibfield  {author} {\bibinfo {author} {\bibfnamefont {M.~W.}\ \bibnamefont
  {Johnson}}, \bibinfo {author} {\bibfnamefont {M.~H.}\ \bibnamefont {Amin}},
  \bibinfo {author} {\bibfnamefont {S.}~\bibnamefont {Gildert}}, \bibinfo
  {author} {\bibfnamefont {T.}~\bibnamefont {Lanting}}, \bibinfo {author}
  {\bibfnamefont {F.}~\bibnamefont {Hamze}}, \bibinfo {author} {\bibfnamefont
  {N.~G.}\ \bibnamefont {Dickson}}, \bibinfo {author} {\bibfnamefont
  {R.}~\bibnamefont {Harris}}, \bibinfo {author} {\bibfnamefont {A.~J.}\
  \bibnamefont {Berkley}}, \bibinfo {author} {\bibfnamefont {J.}~\bibnamefont
  {Johansson}}, \bibinfo {author} {\bibfnamefont {P.~I.}\ \bibnamefont
  {Bunyk}}, \bibinfo {author} {\bibfnamefont {E.~M.}\ \bibnamefont {Chapple}},
  \bibinfo {author} {\bibfnamefont {C.}~\bibnamefont {Enderud}}, \bibinfo
  {author} {\bibfnamefont {J.~P.}\ \bibnamefont {Hilton}}, \bibinfo {author}
  {\bibfnamefont {K.}~\bibnamefont {Karimi}}, \bibinfo {author} {\bibfnamefont
  {E.}~\bibnamefont {Ladizinsky}}, \bibinfo {author} {\bibfnamefont
  {N.}~\bibnamefont {Ladizinsky}}, \bibinfo {author} {\bibfnamefont
  {T.}~\bibnamefont {Oh}}, \bibinfo {author} {\bibfnamefont {I.}~\bibnamefont
  {Perminov}}, \bibinfo {author} {\bibfnamefont {C.}~\bibnamefont {Rich}},
  \bibinfo {author} {\bibfnamefont {M.~C.}\ \bibnamefont {Thom}}, \bibinfo
  {author} {\bibfnamefont {E.}~\bibnamefont {Tolkacheva}}, \bibinfo {author}
  {\bibfnamefont {C.~J.~S.}\ \bibnamefont {Truncik}}, \bibinfo {author}
  {\bibfnamefont {S.}~\bibnamefont {Uchaikin}}, \bibinfo {author}
  {\bibfnamefont {J.}~\bibnamefont {Wang}}, \bibinfo {author} {\bibfnamefont
  {B.}~\bibnamefont {Wilson}}, \ and\ \bibinfo {author} {\bibfnamefont
  {G.}~\bibnamefont {Rose}},\ }\href {\doibase 10.1038/nature10012} {\bibfield
  {journal} {\bibinfo  {journal} {Nature}\ }\textbf {\bibinfo {volume} {473}},\
  \bibinfo {pages} {194} (\bibinfo {year} {2011})}\BibitemShut {NoStop}%
\bibitem [{\citenamefont {Harris}\ \emph {et~al.}(2018)\citenamefont {Harris},
  \citenamefont {Sato}, \citenamefont {Berkley}, \citenamefont {Reis},
  \citenamefont {Altomare}, \citenamefont {Amin}, \citenamefont {Boothby},
  \citenamefont {Bunyk}, \citenamefont {Deng}, \citenamefont {Enderud},
  \citenamefont {Huang}, \citenamefont {Hoskinson}, \citenamefont {Johnson},
  \citenamefont {Ladizinsky}, \citenamefont {Ladizinsky}, \citenamefont
  {Lanting}, \citenamefont {Li}, \citenamefont {Medina}, \citenamefont
  {Molavi}, \citenamefont {Neufeld}, \citenamefont {Oh}, \citenamefont
  {Pavlov}, \citenamefont {Perminov}, \citenamefont {Rich}, \citenamefont
  {Smirnov}, \citenamefont {Swenson}, \citenamefont {Tsai}, \citenamefont
  {Volkmann}, \citenamefont {Whittaker},\ and\ \citenamefont
  {Yao}}]{Harris2018}%
  \BibitemOpen
  \bibfield  {author} {\bibinfo {author} {\bibfnamefont {R.}~\bibnamefont
  {Harris}}, \bibinfo {author} {\bibfnamefont {Y.}~\bibnamefont {Sato}},
  \bibinfo {author} {\bibfnamefont {A.~J.}\ \bibnamefont {Berkley}}, \bibinfo
  {author} {\bibfnamefont {M.}~\bibnamefont {Reis}}, \bibinfo {author}
  {\bibfnamefont {F.}~\bibnamefont {Altomare}}, \bibinfo {author}
  {\bibfnamefont {M.~H.}\ \bibnamefont {Amin}}, \bibinfo {author}
  {\bibfnamefont {K.}~\bibnamefont {Boothby}}, \bibinfo {author} {\bibfnamefont
  {P.~I.}\ \bibnamefont {Bunyk}}, \bibinfo {author} {\bibfnamefont
  {C.}~\bibnamefont {Deng}}, \bibinfo {author} {\bibfnamefont {C.}~\bibnamefont
  {Enderud}}, \bibinfo {author} {\bibfnamefont {S.}~\bibnamefont {Huang}},
  \bibinfo {author} {\bibfnamefont {E.~M.}\ \bibnamefont {Hoskinson}}, \bibinfo
  {author} {\bibfnamefont {M.~W.}\ \bibnamefont {Johnson}}, \bibinfo {author}
  {\bibfnamefont {E.}~\bibnamefont {Ladizinsky}}, \bibinfo {author}
  {\bibfnamefont {N.}~\bibnamefont {Ladizinsky}}, \bibinfo {author}
  {\bibfnamefont {T.}~\bibnamefont {Lanting}}, \bibinfo {author} {\bibfnamefont
  {R.}~\bibnamefont {Li}}, \bibinfo {author} {\bibfnamefont {T.}~\bibnamefont
  {Medina}}, \bibinfo {author} {\bibfnamefont {R.}~\bibnamefont {Molavi}},
  \bibinfo {author} {\bibfnamefont {R.}~\bibnamefont {Neufeld}}, \bibinfo
  {author} {\bibfnamefont {T.}~\bibnamefont {Oh}}, \bibinfo {author}
  {\bibfnamefont {I.}~\bibnamefont {Pavlov}}, \bibinfo {author} {\bibfnamefont
  {I.}~\bibnamefont {Perminov}}, \bibinfo {author} {\bibfnamefont
  {C.}~\bibnamefont {Rich}}, \bibinfo {author} {\bibfnamefont {A.}~\bibnamefont
  {Smirnov}}, \bibinfo {author} {\bibfnamefont {L.}~\bibnamefont {Swenson}},
  \bibinfo {author} {\bibfnamefont {N.}~\bibnamefont {Tsai}}, \bibinfo {author}
  {\bibfnamefont {M.}~\bibnamefont {Volkmann}}, \bibinfo {author}
  {\bibfnamefont {J.}~\bibnamefont {Whittaker}}, \ and\ \bibinfo {author}
  {\bibfnamefont {J.}~\bibnamefont {Yao}},\ }\href {\doibase
  10.1126/science.aat2025} {\bibfield  {journal} {\bibinfo  {journal}
  {Science}\ }\textbf {\bibinfo {volume} {165}},\ \bibinfo {pages} {162}
  (\bibinfo {year} {2018})}\BibitemShut {NoStop}%
\bibitem [{\citenamefont {King}\ \emph {et~al.}(2018)\citenamefont {King},
  \citenamefont {Carrasquilla}, \citenamefont {Raymond}, \citenamefont
  {Ozfidan}, \citenamefont {Andriyash}, \citenamefont {Berkley}, \citenamefont
  {Reis}, \citenamefont {Lanting}, \citenamefont {Harris}, \citenamefont
  {Altomare}, \citenamefont {Boothby}, \citenamefont {Bunyk}, \citenamefont
  {Enderud}, \citenamefont {Fr{\'{e}}chette}, \citenamefont {Hoskinson},
  \citenamefont {Ladizinsky}, \citenamefont {Oh}, \citenamefont
  {Poulin-Lamarre}, \citenamefont {Rich}, \citenamefont {Sato}, \citenamefont
  {Smirnov}, \citenamefont {Swenson}, \citenamefont {Volkmann}, \citenamefont
  {Whittaker}, \citenamefont {Yao}, \citenamefont {Ladizinsky}, \citenamefont
  {Mark}, \citenamefont {Hilton},\ and\ \citenamefont {Amin}}]{King2018}%
  \BibitemOpen
  \bibfield  {author} {\bibinfo {author} {\bibfnamefont {A.~D.}\ \bibnamefont
  {King}}, \bibinfo {author} {\bibfnamefont {J.}~\bibnamefont {Carrasquilla}},
  \bibinfo {author} {\bibfnamefont {J.}~\bibnamefont {Raymond}}, \bibinfo
  {author} {\bibfnamefont {I.}~\bibnamefont {Ozfidan}}, \bibinfo {author}
  {\bibfnamefont {E.}~\bibnamefont {Andriyash}}, \bibinfo {author}
  {\bibfnamefont {A.~J.}\ \bibnamefont {Berkley}}, \bibinfo {author}
  {\bibfnamefont {M.}~\bibnamefont {Reis}}, \bibinfo {author} {\bibfnamefont
  {T.}~\bibnamefont {Lanting}}, \bibinfo {author} {\bibfnamefont
  {R.}~\bibnamefont {Harris}}, \bibinfo {author} {\bibfnamefont
  {F.}~\bibnamefont {Altomare}}, \bibinfo {author} {\bibfnamefont
  {K.}~\bibnamefont {Boothby}}, \bibinfo {author} {\bibfnamefont {P.~I.}\
  \bibnamefont {Bunyk}}, \bibinfo {author} {\bibfnamefont {C.}~\bibnamefont
  {Enderud}}, \bibinfo {author} {\bibfnamefont {A.}~\bibnamefont
  {Fr{\'{e}}chette}}, \bibinfo {author} {\bibfnamefont {E.~M.}\ \bibnamefont
  {Hoskinson}}, \bibinfo {author} {\bibfnamefont {N.}~\bibnamefont
  {Ladizinsky}}, \bibinfo {author} {\bibfnamefont {T.}~\bibnamefont {Oh}},
  \bibinfo {author} {\bibfnamefont {G.}~\bibnamefont {Poulin-Lamarre}},
  \bibinfo {author} {\bibfnamefont {C.}~\bibnamefont {Rich}}, \bibinfo {author}
  {\bibfnamefont {Y.}~\bibnamefont {Sato}}, \bibinfo {author} {\bibfnamefont
  {A.~Y.}\ \bibnamefont {Smirnov}}, \bibinfo {author} {\bibfnamefont {L.~J.}\
  \bibnamefont {Swenson}}, \bibinfo {author} {\bibfnamefont {M.~H.}\
  \bibnamefont {Volkmann}}, \bibinfo {author} {\bibfnamefont {J.}~\bibnamefont
  {Whittaker}}, \bibinfo {author} {\bibfnamefont {J.}~\bibnamefont {Yao}},
  \bibinfo {author} {\bibfnamefont {E.}~\bibnamefont {Ladizinsky}}, \bibinfo
  {author} {\bibfnamefont {W.}~\bibnamefont {Mark}}, \bibinfo {author}
  {\bibfnamefont {J.~P.}\ \bibnamefont {Hilton}}, \ and\ \bibinfo {author}
  {\bibfnamefont {M.~H.}\ \bibnamefont {Amin}},\ }\href {\doibase
  10.1038/s41586-018-0410-x} {\bibfield  {journal} {\bibinfo  {journal}
  {Nature}\ }\textbf {\bibinfo {volume} {560}},\ \bibinfo {pages} {456}
  (\bibinfo {year} {2018})}\BibitemShut {NoStop}%
\bibitem [{\citenamefont {King}\ \emph {et~al.}(2019)\citenamefont {King},
  \citenamefont {Raymond}, \citenamefont {Lanting}, \citenamefont {Isakov},
  \citenamefont {Mohseni}, \citenamefont {Poulin-Lamarre}, \citenamefont
  {Ejtemaee}, \citenamefont {Bernoudy}, \citenamefont {Ozfidan}, \citenamefont
  {Smirnov}, \citenamefont {Reis}, \citenamefont {Altomare}, \citenamefont
  {Babcock}, \citenamefont {Baron}, \citenamefont {Berkley}, \citenamefont
  {Boothby}, \citenamefont {Bunyk}, \citenamefont {Christiani}, \citenamefont
  {Enderud}, \citenamefont {Evert}, \citenamefont {Harris}, \citenamefont
  {Hoskinson}, \citenamefont {Huang}, \citenamefont {Jooya}, \citenamefont
  {Khodabandelou}, \citenamefont {Ladizinsky}, \citenamefont {Li},
  \citenamefont {Lott}, \citenamefont {MacDonald}, \citenamefont {Marsden},
  \citenamefont {Marsden}, \citenamefont {Medina}, \citenamefont {Molavi},
  \citenamefont {Neufeld}, \citenamefont {Norouzpour}, \citenamefont {Oh},
  \citenamefont {Pavlov}, \citenamefont {Perminov}, \citenamefont {Prescott},
  \citenamefont {Rich}, \citenamefont {Sato}, \citenamefont {Sheldan},
  \citenamefont {Sterling}, \citenamefont {Swenson}, \citenamefont {Tsai},
  \citenamefont {Volkmann}, \citenamefont {Whittaker}, \citenamefont
  {Wilkinson}, \citenamefont {Yao}, \citenamefont {Neven}, \citenamefont
  {Hilton}, \citenamefont {Ladizinsky}, \citenamefont {Johnson},\ and\
  \citenamefont {Amin}}]{King2019}%
  \BibitemOpen
  \bibfield  {author} {\bibinfo {author} {\bibfnamefont {A.~D.}\ \bibnamefont
  {King}}, \bibinfo {author} {\bibfnamefont {J.}~\bibnamefont {Raymond}},
  \bibinfo {author} {\bibfnamefont {T.}~\bibnamefont {Lanting}}, \bibinfo
  {author} {\bibfnamefont {S.~V.}\ \bibnamefont {Isakov}}, \bibinfo {author}
  {\bibfnamefont {M.}~\bibnamefont {Mohseni}}, \bibinfo {author} {\bibfnamefont
  {G.}~\bibnamefont {Poulin-Lamarre}}, \bibinfo {author} {\bibfnamefont
  {S.}~\bibnamefont {Ejtemaee}}, \bibinfo {author} {\bibfnamefont
  {W.}~\bibnamefont {Bernoudy}}, \bibinfo {author} {\bibfnamefont
  {I.}~\bibnamefont {Ozfidan}}, \bibinfo {author} {\bibfnamefont {A.~Y.}\
  \bibnamefont {Smirnov}}, \bibinfo {author} {\bibfnamefont {M.}~\bibnamefont
  {Reis}}, \bibinfo {author} {\bibfnamefont {F.}~\bibnamefont {Altomare}},
  \bibinfo {author} {\bibfnamefont {M.}~\bibnamefont {Babcock}}, \bibinfo
  {author} {\bibfnamefont {C.}~\bibnamefont {Baron}}, \bibinfo {author}
  {\bibfnamefont {A.~J.}\ \bibnamefont {Berkley}}, \bibinfo {author}
  {\bibfnamefont {K.}~\bibnamefont {Boothby}}, \bibinfo {author} {\bibfnamefont
  {P.~I.}\ \bibnamefont {Bunyk}}, \bibinfo {author} {\bibfnamefont
  {H.}~\bibnamefont {Christiani}}, \bibinfo {author} {\bibfnamefont
  {C.}~\bibnamefont {Enderud}}, \bibinfo {author} {\bibfnamefont
  {B.}~\bibnamefont {Evert}}, \bibinfo {author} {\bibfnamefont
  {R.}~\bibnamefont {Harris}}, \bibinfo {author} {\bibfnamefont
  {E.}~\bibnamefont {Hoskinson}}, \bibinfo {author} {\bibfnamefont
  {S.}~\bibnamefont {Huang}}, \bibinfo {author} {\bibfnamefont
  {K.}~\bibnamefont {Jooya}}, \bibinfo {author} {\bibfnamefont
  {A.}~\bibnamefont {Khodabandelou}}, \bibinfo {author} {\bibfnamefont
  {N.}~\bibnamefont {Ladizinsky}}, \bibinfo {author} {\bibfnamefont
  {R.}~\bibnamefont {Li}}, \bibinfo {author} {\bibfnamefont {P.~A.}\
  \bibnamefont {Lott}}, \bibinfo {author} {\bibfnamefont {A.~J.~R.}\
  \bibnamefont {MacDonald}}, \bibinfo {author} {\bibfnamefont {D.}~\bibnamefont
  {Marsden}}, \bibinfo {author} {\bibfnamefont {G.}~\bibnamefont {Marsden}},
  \bibinfo {author} {\bibfnamefont {T.}~\bibnamefont {Medina}}, \bibinfo
  {author} {\bibfnamefont {R.}~\bibnamefont {Molavi}}, \bibinfo {author}
  {\bibfnamefont {R.}~\bibnamefont {Neufeld}}, \bibinfo {author} {\bibfnamefont
  {M.}~\bibnamefont {Norouzpour}}, \bibinfo {author} {\bibfnamefont
  {T.}~\bibnamefont {Oh}}, \bibinfo {author} {\bibfnamefont {I.}~\bibnamefont
  {Pavlov}}, \bibinfo {author} {\bibfnamefont {I.}~\bibnamefont {Perminov}},
  \bibinfo {author} {\bibfnamefont {T.}~\bibnamefont {Prescott}}, \bibinfo
  {author} {\bibfnamefont {C.}~\bibnamefont {Rich}}, \bibinfo {author}
  {\bibfnamefont {Y.}~\bibnamefont {Sato}}, \bibinfo {author} {\bibfnamefont
  {B.}~\bibnamefont {Sheldan}}, \bibinfo {author} {\bibfnamefont
  {G.}~\bibnamefont {Sterling}}, \bibinfo {author} {\bibfnamefont {L.~J.}\
  \bibnamefont {Swenson}}, \bibinfo {author} {\bibfnamefont {N.}~\bibnamefont
  {Tsai}}, \bibinfo {author} {\bibfnamefont {M.~H.}\ \bibnamefont {Volkmann}},
  \bibinfo {author} {\bibfnamefont {J.~D.}\ \bibnamefont {Whittaker}}, \bibinfo
  {author} {\bibfnamefont {W.}~\bibnamefont {Wilkinson}}, \bibinfo {author}
  {\bibfnamefont {J.}~\bibnamefont {Yao}}, \bibinfo {author} {\bibfnamefont
  {H.}~\bibnamefont {Neven}}, \bibinfo {author} {\bibfnamefont {J.~P.}\
  \bibnamefont {Hilton}}, \bibinfo {author} {\bibfnamefont {E.}~\bibnamefont
  {Ladizinsky}}, \bibinfo {author} {\bibfnamefont {M.~W.}\ \bibnamefont
  {Johnson}}, \ and\ \bibinfo {author} {\bibfnamefont {M.~H.}\ \bibnamefont
  {Amin}},\ }\href {http://arxiv.org/abs/1911.03446} {\enquote {\bibinfo
  {title} {{Scaling advantage in quantum simulation of geometrically frustrated
  magnets}},}\ } (\bibinfo {year} {2019}),\ \Eprint
  {http://arxiv.org/abs/1911.03446} {arXiv:1911.03446} \BibitemShut {NoStop}%
\bibitem [{\citenamefont {Weinberg}\ \emph {et~al.}(2020)\citenamefont
  {Weinberg}, \citenamefont {Tylutki}, \citenamefont {R{\"{o}}nkk{\"{o}}},
  \citenamefont {Westerholm}, \citenamefont {{\AA}str{\"{o}}m}, \citenamefont
  {Manninen}, \citenamefont {T{\"{o}}rm{\"{a}}},\ and\ \citenamefont
  {Sandvik}}]{Weinberg2019}%
  \BibitemOpen
  \bibfield  {author} {\bibinfo {author} {\bibfnamefont {P.}~\bibnamefont
  {Weinberg}}, \bibinfo {author} {\bibfnamefont {M.}~\bibnamefont {Tylutki}},
  \bibinfo {author} {\bibfnamefont {J.~M.}\ \bibnamefont {R{\"{o}}nkk{\"{o}}}},
  \bibinfo {author} {\bibfnamefont {J.}~\bibnamefont {Westerholm}}, \bibinfo
  {author} {\bibfnamefont {J.~A.}\ \bibnamefont {{\AA}str{\"{o}}m}}, \bibinfo
  {author} {\bibfnamefont {P.}~\bibnamefont {Manninen}}, \bibinfo {author}
  {\bibfnamefont {P.}~\bibnamefont {T{\"{o}}rm{\"{a}}}}, \ and\ \bibinfo
  {author} {\bibfnamefont {A.~W.}\ \bibnamefont {Sandvik}},\ }\href {\doibase
  10.1103/PhysRevLett.124.090502} {\bibfield  {journal} {\bibinfo  {journal}
  {Physical Review Letters}\ }\textbf {\bibinfo {volume} {124}},\ \bibinfo
  {pages} {090502} (\bibinfo {year} {2020})}\BibitemShut {NoStop}%
\bibitem [{\citenamefont {Zhou}\ \emph {et~al.}(2020)\citenamefont {Zhou},
  \citenamefont {Green}, \citenamefont {Dahl},\ and\ \citenamefont
  {Chamon}}]{Zhou2020}%
  \BibitemOpen
  \bibfield  {author} {\bibinfo {author} {\bibfnamefont {S.}~\bibnamefont
  {Zhou}}, \bibinfo {author} {\bibfnamefont {D.}~\bibnamefont {Green}},
  \bibinfo {author} {\bibfnamefont {E.~D.}\ \bibnamefont {Dahl}}, \ and\
  \bibinfo {author} {\bibfnamefont {C.}~\bibnamefont {Chamon}},\ }\href
  {http://arxiv.org/abs/2009.07853} {\enquote {\bibinfo {title} {{Experimental
  Realization of Spin Liquids in a Programmable Quantum Device}},}\ } (\bibinfo
  {year} {2020}),\ \Eprint {http://arxiv.org/abs/2009.07853} {arXiv:2009.07853}
  \BibitemShut {NoStop}%
\bibitem [{\citenamefont {Isakov}\ and\ \citenamefont
  {Moessner}(2003)}]{Isakov2003}%
  \BibitemOpen
  \bibfield  {author} {\bibinfo {author} {\bibfnamefont {S.~V.}\ \bibnamefont
  {Isakov}}\ and\ \bibinfo {author} {\bibfnamefont {R.}~\bibnamefont
  {Moessner}},\ }\href {\doibase 10.1103/PhysRevB.68.104409} {\bibfield
  {journal} {\bibinfo  {journal} {Physical Review B}\ }\textbf {\bibinfo
  {volume} {68}},\ \bibinfo {pages} {104409} (\bibinfo {year}
  {2003})}\BibitemShut {NoStop}%
\bibitem [{\citenamefont {Kageyama}\ \emph
  {et~al.}(1997{\natexlab{a}})\citenamefont {Kageyama}, \citenamefont
  {Yoshimura}, \citenamefont {Kosuge}, \citenamefont {Mitamura},\ and\
  \citenamefont {Goto}}]{Kageyama97a}%
  \BibitemOpen
  \bibfield  {author} {\bibinfo {author} {\bibfnamefont {H.}~\bibnamefont
  {Kageyama}}, \bibinfo {author} {\bibfnamefont {K.}~\bibnamefont {Yoshimura}},
  \bibinfo {author} {\bibfnamefont {K.}~\bibnamefont {Kosuge}}, \bibinfo
  {author} {\bibfnamefont {H.}~\bibnamefont {Mitamura}}, \ and\ \bibinfo
  {author} {\bibfnamefont {T.}~\bibnamefont {Goto}},\ }\href {\doibase
  10.1143/JPSJ.66.1607} {\bibfield  {journal} {\bibinfo  {journal} {Journal of
  the Physical Society of Japan}\ }\textbf {\bibinfo {volume} {66}},\ \bibinfo
  {pages} {1607} (\bibinfo {year} {1997}{\natexlab{a}})}\BibitemShut {NoStop}%
\bibitem [{\citenamefont {Kageyama}\ \emph
  {et~al.}(1997{\natexlab{b}})\citenamefont {Kageyama}, \citenamefont
  {Yoshimura}, \citenamefont {Kosuge}, \citenamefont {Azuma}, \citenamefont
  {Takano}, \citenamefont {Mitamura},\ and\ \citenamefont
  {Goto}}]{Kageyama97b}%
  \BibitemOpen
  \bibfield  {author} {\bibinfo {author} {\bibfnamefont {H.}~\bibnamefont
  {Kageyama}}, \bibinfo {author} {\bibfnamefont {K.}~\bibnamefont {Yoshimura}},
  \bibinfo {author} {\bibfnamefont {K.}~\bibnamefont {Kosuge}}, \bibinfo
  {author} {\bibfnamefont {M.}~\bibnamefont {Azuma}}, \bibinfo {author}
  {\bibfnamefont {M.}~\bibnamefont {Takano}}, \bibinfo {author} {\bibfnamefont
  {H.}~\bibnamefont {Mitamura}}, \ and\ \bibinfo {author} {\bibfnamefont
  {T.}~\bibnamefont {Goto}},\ }\href {\doibase 10.1143/JPSJ.66.3996} {\bibfield
   {journal} {\bibinfo  {journal} {Journal of the Physical Society of Japan}\
  }\textbf {\bibinfo {volume} {66}},\ \bibinfo {pages} {3996} (\bibinfo {year}
  {1997}{\natexlab{b}})}\BibitemShut {NoStop}%
\bibitem [{\citenamefont {Maignan}\ \emph
  {et~al.}(2000{\natexlab{a}})\citenamefont {Maignan}, \citenamefont {Michel},
  \citenamefont {Masset}, \citenamefont {Martin},\ and\ \citenamefont
  {Raveau}}]{Maignan00}%
  \BibitemOpen
  \bibfield  {author} {\bibinfo {author} {\bibfnamefont {A.}~\bibnamefont
  {Maignan}}, \bibinfo {author} {\bibfnamefont {C.}~\bibnamefont {Michel}},
  \bibinfo {author} {\bibfnamefont {A.~C.}\ \bibnamefont {Masset}}, \bibinfo
  {author} {\bibfnamefont {C.}~\bibnamefont {Martin}}, \ and\ \bibinfo {author}
  {\bibfnamefont {B.}~\bibnamefont {Raveau}},\ }\href {\doibase
  10.1007/PL00011051} {\bibfield  {journal} {\bibinfo  {journal} {Eur. Phys. J.
  B}\ }\textbf {\bibinfo {volume} {15}},\ \bibinfo {pages} {657–663}
  (\bibinfo {year} {2000}{\natexlab{a}})}\BibitemShut {NoStop}%
\bibitem [{\citenamefont {Hardy}\ \emph
  {et~al.}(2004{\natexlab{b}})\citenamefont {Hardy}, \citenamefont {Lees},
  \citenamefont {Petrenko}, \citenamefont {Paul}, \citenamefont {Flahaut},
  \citenamefont {H\'ebert},\ and\ \citenamefont {Maignan}}]{Hardy04}%
  \BibitemOpen
  \bibfield  {author} {\bibinfo {author} {\bibfnamefont {V.}~\bibnamefont
  {Hardy}}, \bibinfo {author} {\bibfnamefont {M.~R.}\ \bibnamefont {Lees}},
  \bibinfo {author} {\bibfnamefont {O.~A.}\ \bibnamefont {Petrenko}}, \bibinfo
  {author} {\bibfnamefont {D.~M.}\ \bibnamefont {Paul}}, \bibinfo {author}
  {\bibfnamefont {D.}~\bibnamefont {Flahaut}}, \bibinfo {author} {\bibfnamefont
  {S.}~\bibnamefont {H\'ebert}}, \ and\ \bibinfo {author} {\bibfnamefont
  {A.}~\bibnamefont {Maignan}},\ }\href {\doibase 10.1103/PhysRevB.70.064424}
  {\bibfield  {journal} {\bibinfo  {journal} {Phys. Rev. B}\ }\textbf {\bibinfo
  {volume} {70}},\ \bibinfo {pages} {064424} (\bibinfo {year}
  {2004}{\natexlab{b}})}\BibitemShut {NoStop}%
\bibitem [{\citenamefont {Moyoshi}\ and\ \citenamefont
  {Motoya}(2011)}]{Moyoshi11}%
  \BibitemOpen
  \bibfield  {author} {\bibinfo {author} {\bibfnamefont {T.}~\bibnamefont
  {Moyoshi}}\ and\ \bibinfo {author} {\bibfnamefont {K.}~\bibnamefont
  {Motoya}},\ }\href {\doibase 10.1143/JPSJ.80.034701} {\bibfield  {journal}
  {\bibinfo  {journal} {Journal of the Physical Society of Japan}\ }\textbf
  {\bibinfo {volume} {80}},\ \bibinfo {pages} {034701} (\bibinfo {year}
  {2011})}\BibitemShut {NoStop}%
\bibitem [{\citenamefont {Cevallos}\ \emph {et~al.}(2018)\citenamefont
  {Cevallos}, \citenamefont {Stolze}, \citenamefont {Kong},\ and\ \citenamefont
  {Cava}}]{Cevallos18}%
  \BibitemOpen
  \bibfield  {author} {\bibinfo {author} {\bibfnamefont {F.~A.}\ \bibnamefont
  {Cevallos}}, \bibinfo {author} {\bibfnamefont {K.}~\bibnamefont {Stolze}},
  \bibinfo {author} {\bibfnamefont {T.}~\bibnamefont {Kong}}, \ and\ \bibinfo
  {author} {\bibfnamefont {R.}~\bibnamefont {Cava}},\ }\href {\doibase
  https://doi.org/10.1016/j.materresbull.2018.04.042} {\bibfield  {journal}
  {\bibinfo  {journal} {Materials Research Bulletin}\ }\textbf {\bibinfo
  {volume} {105}},\ \bibinfo {pages} {154 } (\bibinfo {year}
  {2018})}\BibitemShut {NoStop}%
\bibitem [{\citenamefont {Li}\ \emph {et~al.}(2020{\natexlab{a}})\citenamefont
  {Li}, \citenamefont {Liao}, \citenamefont {Chen}, \citenamefont {Zeng},
  \citenamefont {Sheng}, \citenamefont {Qi}, \citenamefont {Meng},\ and\
  \citenamefont {Li}}]{LiHan20}%
  \BibitemOpen
  \bibfield  {author} {\bibinfo {author} {\bibfnamefont {H.}~\bibnamefont
  {Li}}, \bibinfo {author} {\bibfnamefont {Y.~D.}\ \bibnamefont {Liao}},
  \bibinfo {author} {\bibfnamefont {B.-B.}\ \bibnamefont {Chen}}, \bibinfo
  {author} {\bibfnamefont {X.-T.}\ \bibnamefont {Zeng}}, \bibinfo {author}
  {\bibfnamefont {X.-L.}\ \bibnamefont {Sheng}}, \bibinfo {author}
  {\bibfnamefont {Y.}~\bibnamefont {Qi}}, \bibinfo {author} {\bibfnamefont
  {Z.~Y.}\ \bibnamefont {Meng}}, \ and\ \bibinfo {author} {\bibfnamefont
  {W.}~\bibnamefont {Li}},\ }\href {\doibase 10.1038/s41467-020-14907-8}
  {\bibfield  {journal} {\bibinfo  {journal} {Nature Communications}\ }\textbf
  {\bibinfo {volume} {11}},\ \bibinfo {pages} {1111} (\bibinfo {year}
  {2020}{\natexlab{a}})}\BibitemShut {NoStop}%
\bibitem [{\citenamefont {Shen}\ \emph {et~al.}(2019)\citenamefont {Shen},
  \citenamefont {Liu}, \citenamefont {Qin}, \citenamefont {Shen}, \citenamefont
  {Li}, \citenamefont {Bewley}, \citenamefont {Schneidewind}, \citenamefont
  {Chen},\ and\ \citenamefont {Zhao}}]{Shen19}%
  \BibitemOpen
  \bibfield  {author} {\bibinfo {author} {\bibfnamefont {Y.}~\bibnamefont
  {Shen}}, \bibinfo {author} {\bibfnamefont {C.}~\bibnamefont {Liu}}, \bibinfo
  {author} {\bibfnamefont {Y.}~\bibnamefont {Qin}}, \bibinfo {author}
  {\bibfnamefont {S.}~\bibnamefont {Shen}}, \bibinfo {author} {\bibfnamefont
  {Y.-D.}\ \bibnamefont {Li}}, \bibinfo {author} {\bibfnamefont
  {R.}~\bibnamefont {Bewley}}, \bibinfo {author} {\bibfnamefont
  {A.}~\bibnamefont {Schneidewind}}, \bibinfo {author} {\bibfnamefont
  {G.}~\bibnamefont {Chen}}, \ and\ \bibinfo {author} {\bibfnamefont
  {J.}~\bibnamefont {Zhao}},\ }\href {\doibase 10.1038/s41467-019-12410-3}
  {\bibfield  {journal} {\bibinfo  {journal} {Nature Communications}\ }\textbf
  {\bibinfo {volume} {10}},\ \bibinfo {pages} {4530} (\bibinfo {year}
  {2019})}\BibitemShut {NoStop}%
\bibitem [{\citenamefont {Liu}\ \emph {et~al.}(2020)\citenamefont {Liu},
  \citenamefont {Huang},\ and\ \citenamefont {Chen}}]{Liu20}%
  \BibitemOpen
  \bibfield  {author} {\bibinfo {author} {\bibfnamefont {C.}~\bibnamefont
  {Liu}}, \bibinfo {author} {\bibfnamefont {C.-J.}\ \bibnamefont {Huang}}, \
  and\ \bibinfo {author} {\bibfnamefont {G.}~\bibnamefont {Chen}},\ }\href
  {\doibase 10.1103/PhysRevResearch.2.043013} {\bibfield  {journal} {\bibinfo
  {journal} {Physical Review Research}\ }\textbf {\bibinfo {volume} {2}},\
  \bibinfo {pages} {043013} (\bibinfo {year} {2020})}\BibitemShut {NoStop}%
\bibitem [{\citenamefont {Li}\ \emph {et~al.}(2020{\natexlab{b}})\citenamefont
  {Li}, \citenamefont {Bachus}, \citenamefont {Deng}, \citenamefont {Schmidt},
  \citenamefont {Thoma}, \citenamefont {Hutanu}, \citenamefont {Tokiwa},
  \citenamefont {Tsirlin},\ and\ \citenamefont {Gegenwart}}]{LiYueshing20}%
  \BibitemOpen
  \bibfield  {author} {\bibinfo {author} {\bibfnamefont {Y.}~\bibnamefont
  {Li}}, \bibinfo {author} {\bibfnamefont {S.}~\bibnamefont {Bachus}}, \bibinfo
  {author} {\bibfnamefont {H.}~\bibnamefont {Deng}}, \bibinfo {author}
  {\bibfnamefont {W.}~\bibnamefont {Schmidt}}, \bibinfo {author} {\bibfnamefont
  {H.}~\bibnamefont {Thoma}}, \bibinfo {author} {\bibfnamefont
  {V.}~\bibnamefont {Hutanu}}, \bibinfo {author} {\bibfnamefont
  {Y.}~\bibnamefont {Tokiwa}}, \bibinfo {author} {\bibfnamefont {A.~A.}\
  \bibnamefont {Tsirlin}}, \ and\ \bibinfo {author} {\bibfnamefont
  {P.}~\bibnamefont {Gegenwart}},\ }\href {\doibase 10.1103/PhysRevX.10.011007}
  {\bibfield  {journal} {\bibinfo  {journal} {Phys. Rev. X}\ }\textbf {\bibinfo
  {volume} {10}},\ \bibinfo {pages} {011007} (\bibinfo {year}
  {2020}{\natexlab{b}})}\BibitemShut {NoStop}%
\bibitem [{\citenamefont {M{\"{u}}ller-Krumbhaar}\ and\ \citenamefont
  {Binder}(1972)}]{Muller-Krumbhaar1972}%
  \BibitemOpen
  \bibfield  {author} {\bibinfo {author} {\bibfnamefont {H.}~\bibnamefont
  {M{\"{u}}ller-Krumbhaar}}\ and\ \bibinfo {author} {\bibfnamefont
  {K.}~\bibnamefont {Binder}},\ }\href {\doibase 10.1007/BF01379785} {\bibfield
   {journal} {\bibinfo  {journal} {Zeitschrift f{\"{u}}r Physik A Hadrons and
  nuclei}\ }\textbf {\bibinfo {volume} {254}},\ \bibinfo {pages} {269}
  (\bibinfo {year} {1972})}\BibitemShut {NoStop}%
\bibitem [{\citenamefont {Kairys}\ \emph {et~al.}(2020)\citenamefont {Kairys},
  \citenamefont {King}, \citenamefont {Ozfidan}, \citenamefont {Boothby},
  \citenamefont {Raymond}, \citenamefont {Banerjee},\ and\ \citenamefont
  {Humble}}]{Kairys2020}%
  \BibitemOpen
  \bibfield  {author} {\bibinfo {author} {\bibfnamefont {P.}~\bibnamefont
  {Kairys}}, \bibinfo {author} {\bibfnamefont {A.~D.}\ \bibnamefont {King}},
  \bibinfo {author} {\bibfnamefont {I.}~\bibnamefont {Ozfidan}}, \bibinfo
  {author} {\bibfnamefont {K.}~\bibnamefont {Boothby}}, \bibinfo {author}
  {\bibfnamefont {J.}~\bibnamefont {Raymond}}, \bibinfo {author} {\bibfnamefont
  {A.}~\bibnamefont {Banerjee}}, \ and\ \bibinfo {author} {\bibfnamefont
  {T.~S.}\ \bibnamefont {Humble}},\ }\href {\doibase
  10.1103/PRXQuantum.1.020320} {\bibfield  {journal} {\bibinfo  {journal} {PRX
  Quantum}\ }\textbf {\bibinfo {volume} {1}},\ \bibinfo {pages} {020320}
  (\bibinfo {year} {2020})}\BibitemShut {NoStop}%
\bibitem [{\citenamefont {Moessner}\ and\ \citenamefont
  {Sondhi}(2001)}]{Moessner01}%
  \BibitemOpen
  \bibfield  {author} {\bibinfo {author} {\bibfnamefont {R.}~\bibnamefont
  {Moessner}}\ and\ \bibinfo {author} {\bibfnamefont {S.~L.}\ \bibnamefont
  {Sondhi}},\ }\href {\doibase 10.1103/PhysRevB.63.224401} {\bibfield
  {journal} {\bibinfo  {journal} {Phys. Rev. B}\ }\textbf {\bibinfo {volume}
  {63}},\ \bibinfo {pages} {224401} (\bibinfo {year} {2001})}\BibitemShut
  {NoStop}%
\bibitem [{\citenamefont {Kudasov}(2006)}]{Kudasov2006}%
  \BibitemOpen
  \bibfield  {author} {\bibinfo {author} {\bibfnamefont {Y.~B.}\ \bibnamefont
  {Kudasov}},\ }\href {\doibase 10.1103/PhysRevLett.96.027212} {\bibfield
  {journal} {\bibinfo  {journal} {Physical Review Letters}\ }\textbf {\bibinfo
  {volume} {96}},\ \bibinfo {pages} {027212} (\bibinfo {year}
  {2006})}\BibitemShut {NoStop}%
\bibitem [{\citenamefont {Maignan}\ \emph
  {et~al.}(2000{\natexlab{b}})\citenamefont {Maignan}, \citenamefont {Michel},
  \citenamefont {Masset}, \citenamefont {Martin},\ and\ \citenamefont
  {Raveau}}]{Maignan2000}%
  \BibitemOpen
  \bibfield  {author} {\bibinfo {author} {\bibfnamefont {A.}~\bibnamefont
  {Maignan}}, \bibinfo {author} {\bibfnamefont {C.}~\bibnamefont {Michel}},
  \bibinfo {author} {\bibfnamefont {A.~C.}\ \bibnamefont {Masset}}, \bibinfo
  {author} {\bibfnamefont {C.}~\bibnamefont {Martin}}, \ and\ \bibinfo {author}
  {\bibfnamefont {B.}~\bibnamefont {Raveau}},\ }\href {\doibase
  10.1007/PL00011051} {\bibfield  {journal} {\bibinfo  {journal} {European
  Physical Journal B}\ }\textbf {\bibinfo {volume} {15}},\ \bibinfo {pages}
  {657} (\bibinfo {year} {2000}{\natexlab{b}})}\BibitemShut {NoStop}%
\bibitem [{\citenamefont {Boothby}\ \emph {et~al.}(2020)\citenamefont
  {Boothby}, \citenamefont {Bunyk}, \citenamefont {Raymond},\ and\
  \citenamefont {Roy}}]{Boothby2020}%
  \BibitemOpen
  \bibfield  {author} {\bibinfo {author} {\bibfnamefont {K.}~\bibnamefont
  {Boothby}}, \bibinfo {author} {\bibfnamefont {P.}~\bibnamefont {Bunyk}},
  \bibinfo {author} {\bibfnamefont {J.}~\bibnamefont {Raymond}}, \ and\
  \bibinfo {author} {\bibfnamefont {A.}~\bibnamefont {Roy}},\ }\href
  {http://arxiv.org/abs/2003.00133} {\enquote {\bibinfo {title}
  {{Next-Generation Topology of D-Wave Quantum Processors}},}\ } (\bibinfo
  {year} {2020}),\ \Eprint {http://arxiv.org/abs/2003.00133} {arXiv:2003.00133}
  \BibitemShut {NoStop}%
\bibitem [{\citenamefont {Andriyash}\ and\ \citenamefont
  {Amin}(2017)}]{Andriyash2017}%
  \BibitemOpen
  \bibfield  {author} {\bibinfo {author} {\bibfnamefont {E.}~\bibnamefont
  {Andriyash}}\ and\ \bibinfo {author} {\bibfnamefont {M.~H.}\ \bibnamefont
  {Amin}},\ }\href {http://arxiv.org/abs/1703.09277} {\enquote {\bibinfo
  {title} {{Can quantum Monte Carlo simulate quantum annealing?}}}\ } (\bibinfo
  {year} {2017}),\ \Eprint {http://arxiv.org/abs/1703.09277} {arXiv:1703.09277}
  \BibitemShut {NoStop}%
\bibitem [{\citenamefont {Kechedzhi}\ \emph {et~al.}(2018)\citenamefont
  {Kechedzhi}, \citenamefont {Smelyanskiy}, \citenamefont {McClean},
  \citenamefont {Denchev}, \citenamefont {Mohseni}, \citenamefont {Isakov},
  \citenamefont {Boixo}, \citenamefont {Altshuler},\ and\ \citenamefont
  {Neven}}]{Kechedzhi2018}%
  \BibitemOpen
  \bibfield  {author} {\bibinfo {author} {\bibfnamefont {K.}~\bibnamefont
  {Kechedzhi}}, \bibinfo {author} {\bibfnamefont {V.}~\bibnamefont
  {Smelyanskiy}}, \bibinfo {author} {\bibfnamefont {J.~R.}\ \bibnamefont
  {McClean}}, \bibinfo {author} {\bibfnamefont {V.~S.}\ \bibnamefont
  {Denchev}}, \bibinfo {author} {\bibfnamefont {M.}~\bibnamefont {Mohseni}},
  \bibinfo {author} {\bibfnamefont {S.}~\bibnamefont {Isakov}}, \bibinfo
  {author} {\bibfnamefont {S.}~\bibnamefont {Boixo}}, \bibinfo {author}
  {\bibfnamefont {B.}~\bibnamefont {Altshuler}}, \ and\ \bibinfo {author}
  {\bibfnamefont {H.}~\bibnamefont {Neven}},\ }in\ \href {\doibase
  10.4230/LIPIcs.TQC.2018.9} {\emph {\bibinfo {booktitle} {13th Conference on
  the Theory of Quantum Computation, Communication and Cryptography (TQC
  2018)}}},\ \bibinfo {series} {Leibniz International Proceedings in
  Informatics (LIPIcs)}, Vol.\ \bibinfo {volume} {111},\ \bibinfo {editor}
  {edited by\ \bibinfo {editor} {\bibfnamefont {S.}~\bibnamefont {Jeffery}}}\
  (\bibinfo  {publisher} {Schloss Dagstuhl--Leibniz-Zentrum fuer Informatik},\
  \bibinfo {address} {Dagstuhl, Germany},\ \bibinfo {year} {2018})\ pp.\
  \bibinfo {pages} {9:1----9:16}\BibitemShut {NoStop}%
\bibitem [{\citenamefont {Ozfidan}\ \emph {et~al.}(2020)\citenamefont
  {Ozfidan}, \citenamefont {Deng}, \citenamefont {Smirnov}, \citenamefont
  {Lanting}, \citenamefont {Harris}, \citenamefont {Swenson}, \citenamefont
  {Whittaker}, \citenamefont {Altomare}, \citenamefont {Babcock}, \citenamefont
  {Baron}, \citenamefont {Berkley}, \citenamefont {Boothby}, \citenamefont
  {Christiani}, \citenamefont {Bunyk}, \citenamefont {Enderud}, \citenamefont
  {Evert}, \citenamefont {Hager}, \citenamefont {Hajda}, \citenamefont
  {Hilton}, \citenamefont {Huang}, \citenamefont {Hoskinson}, \citenamefont
  {Johnson}, \citenamefont {Jooya}, \citenamefont {Ladizinsky}, \citenamefont
  {Ladizinsky}, \citenamefont {Li}, \citenamefont {MacDonald}, \citenamefont
  {Marsden}, \citenamefont {Marsden}, \citenamefont {Medina}, \citenamefont
  {Molavi}, \citenamefont {Neufeld}, \citenamefont {Nissen}, \citenamefont
  {Norouzpour}, \citenamefont {Oh}, \citenamefont {Pavlov}, \citenamefont
  {Perminov}, \citenamefont {Poulin-Lamarre}, \citenamefont {Reis},
  \citenamefont {Prescott}, \citenamefont {Rich}, \citenamefont {Sato},
  \citenamefont {Sterling}, \citenamefont {Tsai}, \citenamefont {Volkmann},
  \citenamefont {Wilkinson}, \citenamefont {Yao},\ and\ \citenamefont
  {Amin}}]{Ozfidan2019}%
  \BibitemOpen
  \bibfield  {author} {\bibinfo {author} {\bibfnamefont {I.}~\bibnamefont
  {Ozfidan}}, \bibinfo {author} {\bibfnamefont {C.}~\bibnamefont {Deng}},
  \bibinfo {author} {\bibfnamefont {A.}~\bibnamefont {Smirnov}}, \bibinfo
  {author} {\bibfnamefont {T.}~\bibnamefont {Lanting}}, \bibinfo {author}
  {\bibfnamefont {R.}~\bibnamefont {Harris}}, \bibinfo {author} {\bibfnamefont
  {L.}~\bibnamefont {Swenson}}, \bibinfo {author} {\bibfnamefont
  {J.}~\bibnamefont {Whittaker}}, \bibinfo {author} {\bibfnamefont
  {F.}~\bibnamefont {Altomare}}, \bibinfo {author} {\bibfnamefont
  {M.}~\bibnamefont {Babcock}}, \bibinfo {author} {\bibfnamefont
  {C.}~\bibnamefont {Baron}}, \bibinfo {author} {\bibfnamefont
  {A.}~\bibnamefont {Berkley}}, \bibinfo {author} {\bibfnamefont
  {K.}~\bibnamefont {Boothby}}, \bibinfo {author} {\bibfnamefont
  {H.}~\bibnamefont {Christiani}}, \bibinfo {author} {\bibfnamefont
  {P.}~\bibnamefont {Bunyk}}, \bibinfo {author} {\bibfnamefont
  {C.}~\bibnamefont {Enderud}}, \bibinfo {author} {\bibfnamefont
  {B.}~\bibnamefont {Evert}}, \bibinfo {author} {\bibfnamefont
  {M.}~\bibnamefont {Hager}}, \bibinfo {author} {\bibfnamefont
  {A.}~\bibnamefont {Hajda}}, \bibinfo {author} {\bibfnamefont
  {J.}~\bibnamefont {Hilton}}, \bibinfo {author} {\bibfnamefont
  {S.}~\bibnamefont {Huang}}, \bibinfo {author} {\bibfnamefont
  {E.}~\bibnamefont {Hoskinson}}, \bibinfo {author} {\bibfnamefont
  {M.}~\bibnamefont {Johnson}}, \bibinfo {author} {\bibfnamefont
  {K.}~\bibnamefont {Jooya}}, \bibinfo {author} {\bibfnamefont
  {E.}~\bibnamefont {Ladizinsky}}, \bibinfo {author} {\bibfnamefont
  {N.}~\bibnamefont {Ladizinsky}}, \bibinfo {author} {\bibfnamefont
  {R.}~\bibnamefont {Li}}, \bibinfo {author} {\bibfnamefont {A.}~\bibnamefont
  {MacDonald}}, \bibinfo {author} {\bibfnamefont {D.}~\bibnamefont {Marsden}},
  \bibinfo {author} {\bibfnamefont {G.}~\bibnamefont {Marsden}}, \bibinfo
  {author} {\bibfnamefont {T.}~\bibnamefont {Medina}}, \bibinfo {author}
  {\bibfnamefont {R.}~\bibnamefont {Molavi}}, \bibinfo {author} {\bibfnamefont
  {R.}~\bibnamefont {Neufeld}}, \bibinfo {author} {\bibfnamefont
  {M.}~\bibnamefont {Nissen}}, \bibinfo {author} {\bibfnamefont
  {M.}~\bibnamefont {Norouzpour}}, \bibinfo {author} {\bibfnamefont
  {T.}~\bibnamefont {Oh}}, \bibinfo {author} {\bibfnamefont {I.}~\bibnamefont
  {Pavlov}}, \bibinfo {author} {\bibfnamefont {I.}~\bibnamefont {Perminov}},
  \bibinfo {author} {\bibfnamefont {G.}~\bibnamefont {Poulin-Lamarre}},
  \bibinfo {author} {\bibfnamefont {M.}~\bibnamefont {Reis}}, \bibinfo {author}
  {\bibfnamefont {T.}~\bibnamefont {Prescott}}, \bibinfo {author}
  {\bibfnamefont {C.}~\bibnamefont {Rich}}, \bibinfo {author} {\bibfnamefont
  {Y.}~\bibnamefont {Sato}}, \bibinfo {author} {\bibfnamefont {G.}~\bibnamefont
  {Sterling}}, \bibinfo {author} {\bibfnamefont {N.}~\bibnamefont {Tsai}},
  \bibinfo {author} {\bibfnamefont {M.}~\bibnamefont {Volkmann}}, \bibinfo
  {author} {\bibfnamefont {W.}~\bibnamefont {Wilkinson}}, \bibinfo {author}
  {\bibfnamefont {J.}~\bibnamefont {Yao}}, \ and\ \bibinfo {author}
  {\bibfnamefont {M.}~\bibnamefont {Amin}},\ }\href {\doibase
  10.1103/PhysRevApplied.13.034037} {\bibfield  {journal} {\bibinfo  {journal}
  {Physical Review Applied}\ }\textbf {\bibinfo {volume} {13}},\ \bibinfo
  {pages} {034037} (\bibinfo {year} {2020})}\BibitemShut {NoStop}%
\bibitem [{\citenamefont {Nishimura}\ \emph {et~al.}(2020)\citenamefont
  {Nishimura}, \citenamefont {Nishimori},\ and\ \citenamefont
  {Katzgraber}}]{Nishimura2020}%
  \BibitemOpen
  \bibfield  {author} {\bibinfo {author} {\bibfnamefont {K.}~\bibnamefont
  {Nishimura}}, \bibinfo {author} {\bibfnamefont {H.}~\bibnamefont
  {Nishimori}}, \ and\ \bibinfo {author} {\bibfnamefont {H.~G.}\ \bibnamefont
  {Katzgraber}},\ }\href {\doibase 10.1103/PhysRevA.102.042403} {\bibfield
  {journal} {\bibinfo  {journal} {Physical Review A}\ }\textbf {\bibinfo
  {volume} {102}},\ \bibinfo {pages} {042403} (\bibinfo {year}
  {2020})}\BibitemShut {NoStop}%
\bibitem [{\citenamefont {Isakov}\ \emph {et~al.}(2016)\citenamefont {Isakov},
  \citenamefont {Mazzola}, \citenamefont {Smelyanskiy}, \citenamefont {Jiang},
  \citenamefont {Boixo}, \citenamefont {Neven},\ and\ \citenamefont
  {Troyer}}]{Isakov2016}%
  \BibitemOpen
  \bibfield  {author} {\bibinfo {author} {\bibfnamefont {S.~V.}\ \bibnamefont
  {Isakov}}, \bibinfo {author} {\bibfnamefont {G.}~\bibnamefont {Mazzola}},
  \bibinfo {author} {\bibfnamefont {V.~N.}\ \bibnamefont {Smelyanskiy}},
  \bibinfo {author} {\bibfnamefont {Z.}~\bibnamefont {Jiang}}, \bibinfo
  {author} {\bibfnamefont {S.}~\bibnamefont {Boixo}}, \bibinfo {author}
  {\bibfnamefont {H.}~\bibnamefont {Neven}}, \ and\ \bibinfo {author}
  {\bibfnamefont {M.}~\bibnamefont {Troyer}},\ }\href {\doibase
  10.1103/PhysRevLett.117.180402} {\bibfield  {journal} {\bibinfo  {journal}
  {Physical Review Letters}\ }\textbf {\bibinfo {volume} {117}},\ \bibinfo
  {pages} {180402} (\bibinfo {year} {2016})}\BibitemShut {NoStop}%
\end{thebibliography}%
\appendix

\begin{figure*}
  \includegraphics[scale = 0.8]{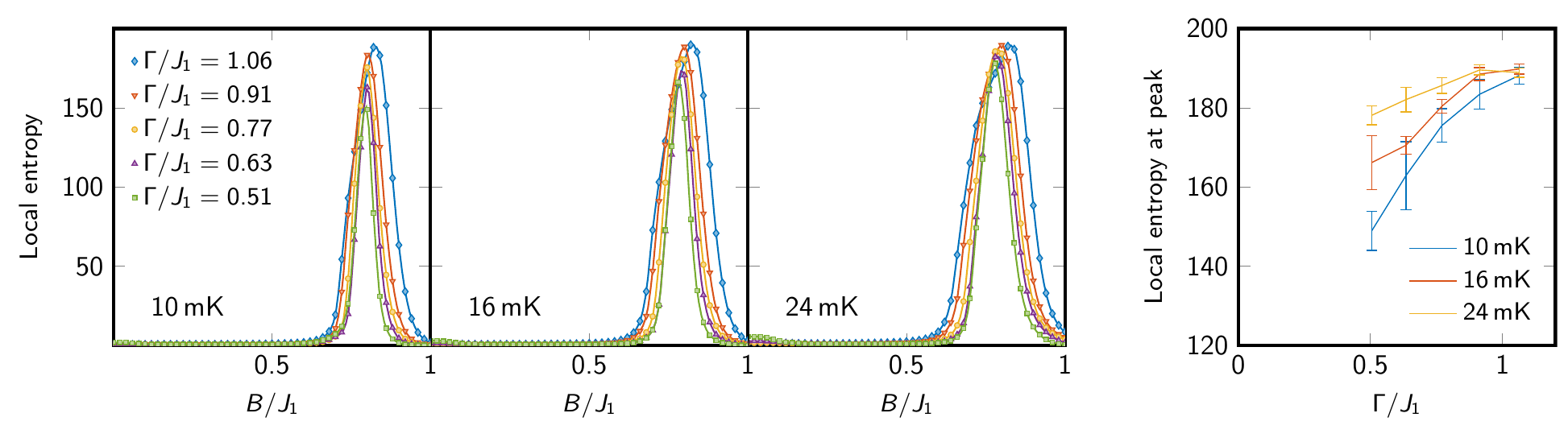}
  \caption{{\bf Local entropy of observed states.} The ground state at $\Gamma=0$, $T=0$, $H=\tfrac 32$ has extensive entropy, consisting of triangular plaquettes with one or zero down-spins, which is lifted by the transverse field.  Local entropy, which we define as the number of neighboring states differing by a swap of a single up-down spin chain pair, becomes increasingly favorable with increasing $\Gamma$ and $T$.  Error bars indicate standard deviation of estimates for individual 50-sample QA programmings.}\label{fig:shoulder}
\end{figure*}
\section{Methods}

\begin{figure}
  \includegraphics[scale = 0.7]{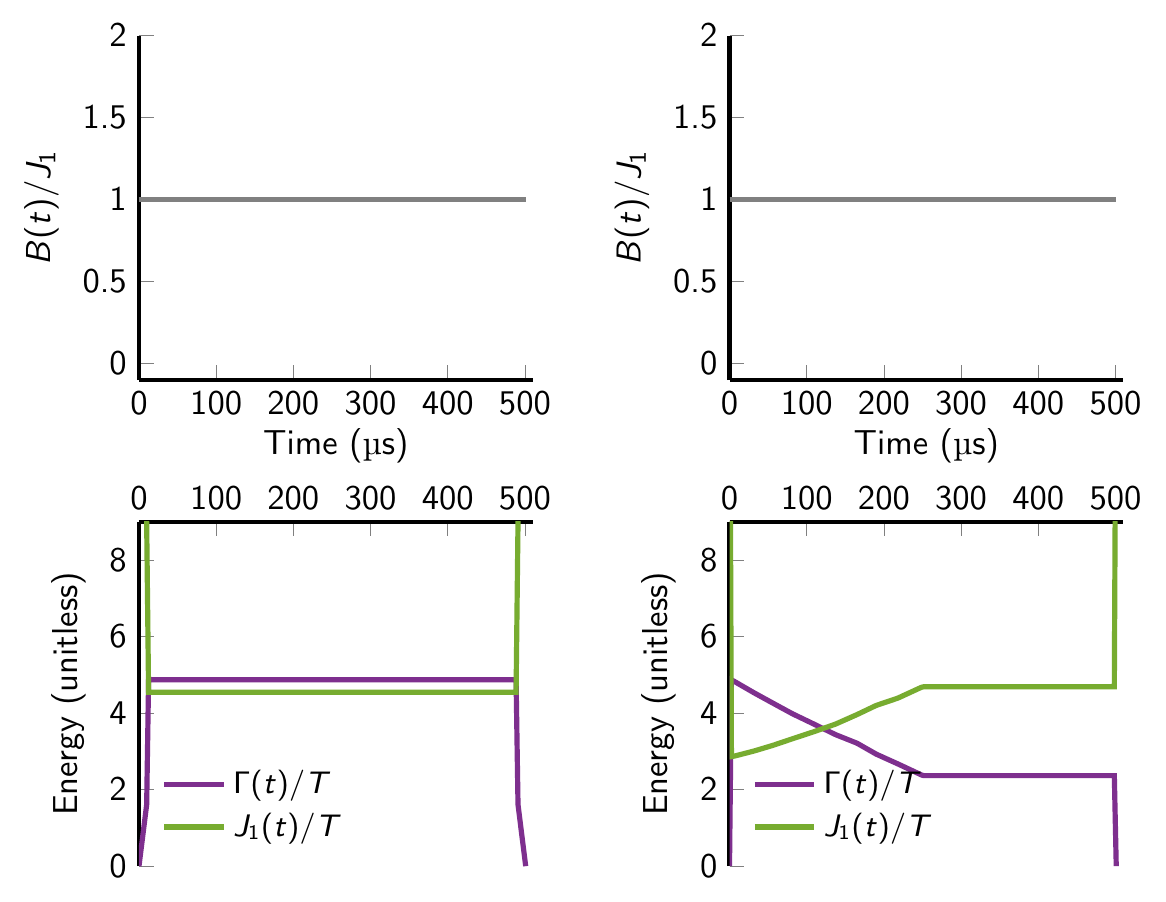}
  \caption{{\bf Reverse anneal protocols for equilibrium estimates.} For estimating equilibrium statistics, we perform ``quantum evolution Monte Carlo'', forming a chain of reverse anneals.  Each reverse anneal starts from a classical state with $\Gamma=0$, then increases $\Gamma$, dwells, and decreases $\Gamma$ to $0$, arriving at a new classical state.  In the easy case (left) the protocol involves a simple dwell.  In the slow-relaxing case (right) $\Gamma$ is attenuated from a higher value before the dwell.  Each chain repeats the reverse anneal 100 times, producing 100 classical output samples that ideally represent projections of the system to the $\sigma^z$ basis.  In both cases, $B(t)/J_1$ is held constant (shown here for example $B(t)/J_1=1$), in contrast to the out-of-equilibrium protocols shown in Fig.~\ref{fig:2}.}\label{fig:qemc}
\end{figure}

\begin{figure*}
  \includegraphics[scale = 0.65]{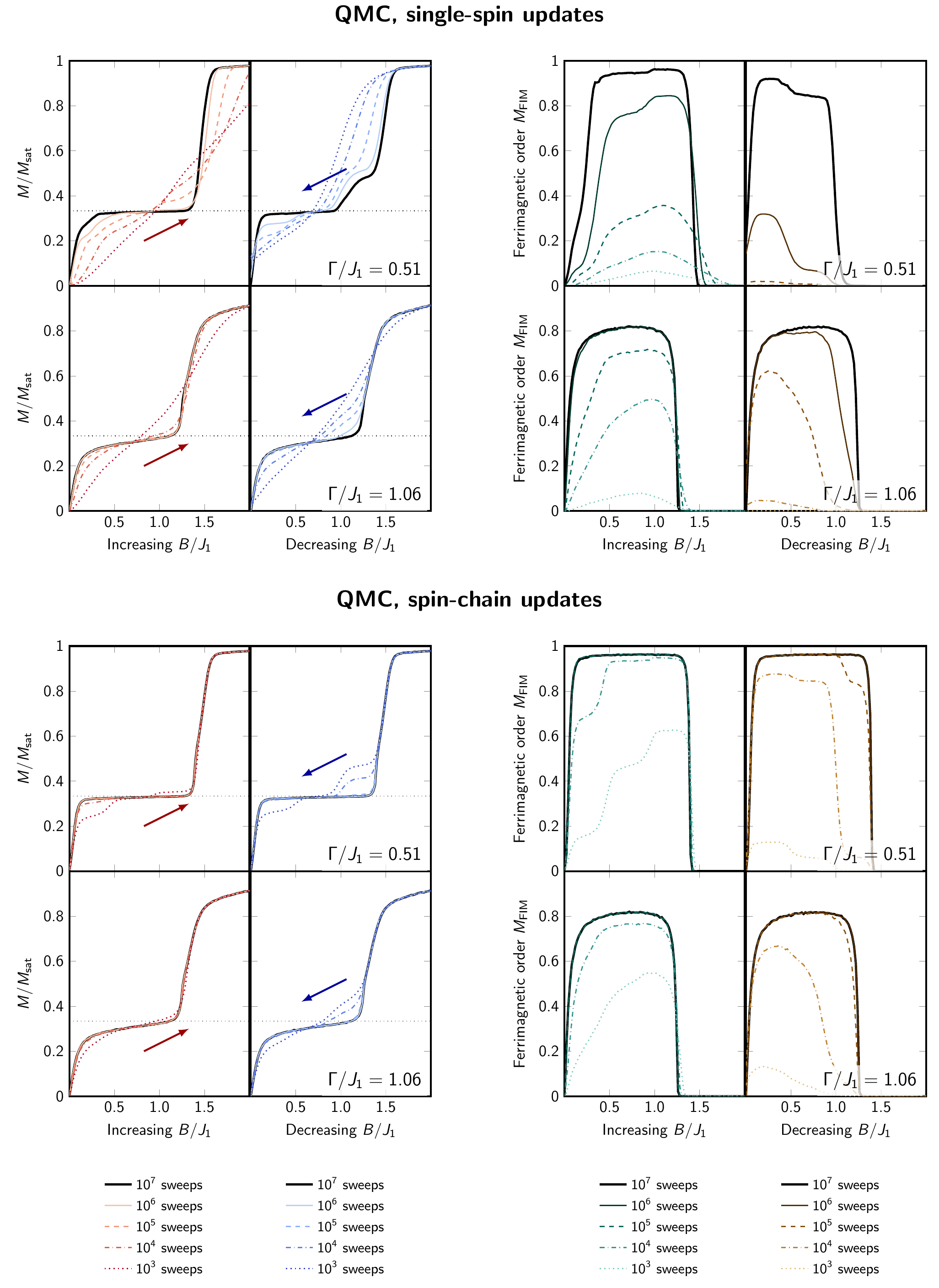}
  \caption{{\bf Out-of-equilibrium QMC simulations.} Simulations are performed on a fully-periodic lattice with no site vacancies, with a longitudinal field swept from $10^3$ to $10^7$ Monte Carlo sweeps.  Sweeps are performed either one spin at a time (top) or four spins at a time (bottom).  Compare with Fig.~\ref{fig:2}.}\label{fig:qmc}
\end{figure*}

\subsection{Experimental methods}

The QA processor used was a D-Wave 2000Q system, in which 2041 of 2048 superconducting rf-SQUID flux qubits were operable.  Effective qubit temperatures of between $\SI{10}{mK}$ and  $\SI{24}{mK}$ were measured via qubit susceptibility following the methods of \cite{Johnson2011} (SM page 8).  Effective TFIM parameters $\Gamma$ and $J_1$ were extracted using a combination of spectroscopy measurements and diagonalization of the Hamiltonian of 12 qubits arranged in a representative gadget, as detailed in the Methods section of \cite{King2019} for a different superconducting processor of similar design.  Similarly, the leading qubit-to-spin nonideality of background susceptibility is compensated in the coupling energies as in Methods of \cite{King2019}.

In this work we study a single lattice programmed into the qubit architecture; this has the same 1800-qubit size as studied in \cite{King2018} but due to inoperable qubits only uses 1764 spins.  Longitudinal fields $H=B/J_1$ can be programmed either negative or positive; we run experiments for $H = 0$, $0.04$, $\ldots$, $2$ in both directions and combine symmetric results.

With each programming of the quantum processing unit (QPU) we draw 100 samples.  Out-of-equilibrium experiments are performed using the protocol described in the main text and shown in Fig.~\ref{fig:2}; these protocols require the ``$H$-gain'' control.  Equilibrium estimates are generated using quantum evolution Monte Carlo (QEMC) \cite{King2018,King2019,Kairys2020}, a method that, analogous to Markov-chain Monte Carlo, forms a Monte Carlo chain of reverse annealing protocols that dwell at the desired Hamiltonian rather than annealing to it.  The QEMC (equilibrium estimating) calls also draw 100 samples, of which we discard the first 50 as QEMC burn-in.We dwell for $\SI{500}{\micro s}$ per step, except in the case where $\Gamma/J_1 < 0.77$, where relaxation is slow.  In this case we reverse anneal to an intermediate value of $\Gamma/J_1$, then anneal ``forward'' (i.e., attenuating $\Gamma$) to the target value of $\Gamma/J_1$ over $\SI{250}{\micro s}$, then dwell at the target $\Gamma/J_1$ for the remaining $\SI{250}{\micro s}$.  Fig.~\ref{fig:qemc} shows the waveforms for a single reverse annealing step in each of these cases.

In practice the relative strengths of $\Gamma$ and $J_1$ are not controlled independently, but rather through single annealing parameter $s$; the functions $\Gamma(s)$ and $J_1(s)$ are fixed functions of the QA processor, which allows time-dependent control of $s(t)$ (cf.~\cite{King2019} Methods).  Varying $\Gamma$ for fixed $J_1$ therefore requires tuning $\Gamma(s)$ and $J_1(s)$ via $s$, then compensating by adjusting the individual programmed coupling terms.

All reported data are taken over $120$ individual programmings, representing $12,000$ samples for out-of-equilibrium and $6,000$ samples for equilibrium.  Reported bulk averages (magnetization, susceptibility, FIM order) are taken over non-boundary spins to reduce the effect of open boundaries and defects.

\subsection{Calibration refinement}

QA simulations in highly degenerate spaces have shown the importance of fine-tuning Hamiltonian parameters calibration to compensate for device variation and boundary conditions \cite{King2018,King2019,Kairys2020,Nishimura2020} (specifically cf.~\cite{King2018} Extended Data Fig.~7).  The idea of per-spin tuning of longitudinal fields in the presence of open boundaries goes back to the 1970s \cite{Muller-Krumbhaar1972}, and although it now seems obsolete for computer simulations, here it allows us to better simulate the thermodynamic limit of a large system in a uniform magnetic field \cite{Kairys2020}, in spite of the fact that limited qubit connectivity prevents us from implementing fully periodic boundaries.

In contrast to previous studies on cylindrical lattices \cite{King2018,King2019} that attempted to faithfully simulate a finite system with partially open boundaries, here we are interested in the thermodynamic limit.  We therefore use the same approach of iteratively tuning $J_{ij}$ terms to homogenize spin-spin correlations.  However, instead of doing this over small rotational symmetry groups of the cylinder, we homogenize correlations and magnetizations across the entire lattice.

When homogenizing the coupling terms we tune terms using $H=B/J_1=0$ observations and propagate the couplings to all simulated $H$ values.  When homogenizing the per-spin field strengths we use all observations and employ smoothing of the Hamiltonian terms so that relative field strengths vary smoothly as a function of $H$.

\section{Entropic origin of magnetization shoulder}

Further evidence for the entropic origin of the shoulder is shown in Fig.~\ref{fig:shoulder}, where we compute the average local degeneracy of observed states, defined as the number of states reachable from a given state by moving a single down-spin chain to a neighboring site.  This local entropy is found to peak at the phase transition, with peak height that increases with both quantum and thermal fluctuations, in the form of changing $\Gamma$ and $T$ respectively.

\section{QMC simulations}

Path-integral Monte Carlo is a common and accurate tool for estimating equilibrium statistics of TFIM systems at finite temperature.  Although in some incoherent cases the dynamics can be directly mapped to tunneling in a quantum system \cite{Isakov2016}, this is not true in general \cite{Andriyash2017,King2019}.  Therefore QA, which involves relaxation of an engineered quantum system, may eventually provide more faithful simulation of the relaxation dynamics of effective transverse-field Ising models such as TMGO.  

Here we draw compare and contrast the out-of-equilibrium QA results with qualitatively analogous QMC simulations.  Fig.~\ref{fig:qmc} shows results on a 1764-spin lattice ($21\times 21$ four-spin chains) with fully periodic boundary conditions.  We consider two QMC variants: one that uses single-spin updates, and one that collectively updates the four-spin chains.  Both are detailed in the supplemental material of Ref.~\cite{King2019}.  We emphasize that these QMC experiments are on a different lattice than the QA experiments, and are only intended to provide qualitative comparison.  Unlike Ref.~\cite{King2019}, we are not attempting a quantitative comparison of relaxation dynamics.

Most obviously, and particularly for large $\Gamma$, the QMC results differ from QA in that they do not saturate either the FIM LRO or the magnetization.  This is because of canting in the $\sigma^x$ direction due to the transverse field; this canting is destroyed as a local relaxation during the QA readout quench.  In one sense this is a drawback of the current QA methods, albeit one that is easily recovered with postprocessing.  A preferable alternative would be the development of faster quench in QA, which can faithfully project to the $\sigma^z$ basis without distortion.  In another sense, however, the method of periodically driving toward a classical ground state is a potentially powerful computational method.

We note that with spin-chain updates, the larger transverse field has little effect on the hysteresis; this is not surprising since the spin-chain updates effectively reduce the system to a two-dimensional triangular AFM.  With single-spin updates, the transverse field indeed accelerates the generation of FIM long-range order as it does in QA.  This is expected, since the origin of the four-spin update is not physical but rather algorithmic.  We expect that future experiments on systems with longer spin chains will elucidate the comparison between QMC and QA in this context.

\newpage

\clearpage
\end{document}